\def\year{2021}\relax
\documentclass[letterpaper]{article} 

\usepackage[hyphens]{url}  
\usepackage{natbib}  
\usepackage{caption} 
\usepackage{aaai21}
\usepackage{times}
\usepackage{helvet}
\usepackage{courier}
\usepackage{booktabs} 
\usepackage{subfigure}
\usepackage{graphicx}
\usepackage{amsmath}
\usepackage{amssymb}
\usepackage{xcolor}
\usepackage{multirow}
\usepackage{balance}
\usepackage[switch]{lineno}

\nocopyright
\begin{document}
\title{Disruption in the Chinese E-Commerce During COVID-19}

\author {
    Yuan Yuan,\textsuperscript{\rm 1}
    Muzhi Guan,\textsuperscript{\rm 1}
    Zhilun Zhou,\textsuperscript{\rm 1}
    Sundong Kim,\textsuperscript{\rm 2}
    Meeyoung Cha,\textsuperscript{\rm 2}
    Depeng Jin,\textsuperscript{\rm 1}
    Yong Li\textsuperscript{\rm 1}
    \\
}
\affiliations {
    \textsuperscript{\rm 1} Tsinghua University, China \\
    \textsuperscript{\rm 2} Institute for Basic Science, South Korea \\
}

\maketitle

\begin{abstract}

The recent outbreak of the novel coronavirus (COVID-19) has infected millions of citizens worldwide and claimed many lives. This paper examines its impact on the Chinese e-commerce market by analyzing behavioral changes seen from a large online shopping platform. We first conduct a time series analysis to identify product categories that faced the most extensive disruptions. The time-lagged analysis shows that behavioral patterns seen in shopping actions are highly responsive to epidemic development. Based on these findings, we present a consumer demand prediction method by encompassing the epidemic statistics and behavioral features for COVID-19 related products. Experiment results demonstrate that our predictions outperform existing baselines and further extend to the long-term and province-level forecasts. We discuss how our market analysis and prediction can help better prepare for future pandemics by gaining an extra time to launch preventive steps.

\end{abstract}

\section{Introduction}

The coronavirus disease 2019 (COVID-19) had a massive breakout in Wuhan during the Spring Festival in 2020 and later followed by a planetary health emergency.  The novel virus had significantly influenced people's daily lives. Governments and the World Health Organization (WHO) have recommended people to stay at home and avoid crowded places. The disease dynamic was particularly rapid in China; it spread from Wuhan to all other regions and became nearly contained over the span of only two months.

According to the McKinsey report~\cite{Mckinsey_global}, global citizens increased their reliance on online shopping and delivery of essential goods, compared to the pre-pandemic time. Under this circumstance, epidemic-related products such as face masks and disinfectants were short in supply, failing to meet people's demand. Such a disruption in the supply-and-demand could reshape online shopping patterns, not only for COVID-19 related products but also for ordinary products. Understanding disruption in e-commerce could benefit all stakeholders (i.e., retailers, consumers, suppliers, delivery systems, and local governments) better prepare for the next pandemic. However, due to the lack of data or scenarios of a time-concentrated outbreak, little is known about consumer demand in the context of epidemics. 

This paper conducts an extensive analysis of online shopping trends before and during the COVID-19 epidemic from the view of a popular e-commerce platform in China, Beidian\footnote{www.beidian.com}~\cite{cao2020beidian}. We characterize the pandemic's impact on the market from changes in product-level demand and supply. First, our analysis reveals which products increased or decreased in sales (after discounting seasonal variation), helping understand how households are coping with the pandemic. Second, it also compares the differences in browsing, searching, and purchasing activities towards pandemic-related goods such as face masks, disinfectants, and thermometers, which reveals the intricate relationship between supply and demand. We also identify which products decreased the largest in sales to infer the causes of such a drop.  We also use time-lagged cross-correlation to quantify how shopping actions respond to the pandemic, interrogating product supply shortage. To the best of our knowledge, no other research has examined the disruptions in e-commerce at such a fine-grained level.

Based on these observations, we propose an \textbf{\underline{En}}coder-decoder model that leverages the online shopping behaviors and the \textbf{\underline{CO}}VID-19 epidemic statistics to predict changes in \textbf{\underline{d}}emand of critical goods, \textbf{EnCod} for short. Experiments show that both shopping and epidemic features gathered from the past weeks are important for predicting pandemic-relevant goods in the upcoming days. Our model achieves higher prediction performance than baselines, and it could be fine-tuned at the province level; each province or city may adopt our model to understand the needs of their citizens during a pandemic. In summary, our main contributions are as follows: \looseness=-1
\begin{enumerate}
    \item We operationalize and release a dataset\footnote{https://bit.ly/3kcAEN5} of people's online shopping behaviors during the COVID-19 epidemic, and its multiple features characterize the marketing change during this period. \looseness=-1
    \item We investigate the changes of different online shopping behaviors, including purchasing, browsing, and searching on the platform, and examine the interplay between the COVID-19 epidemic and consumers' behaviors.
    \item We conduct a time-lagged cross-correlation analysis could reveal which products exhibit a demand pattern that coaligns well with the epidemic dynamics.
    \item We propose a model to forecast consumer demand on essential product categories, and demonstrate our model's effectiveness with regional and long-term forecast.
\end{enumerate}

\section{Related Work}

Various studies have appeared since the outbreak of COVID-19 due to its unprecedented challenges for industry and society. Researchers have studied the impact of epidemics from multiple aspects, including transportation~\cite{huang2020understanding}, gender equality~\cite{alon2020impact}, global poverty~\cite{sumner2020estimates}, and stock market~\cite{baker2020unprecedented}. Among those related to businesses, one study examines the online shopping food services during the government's stay-at-home order~\cite{chang2020covid}. Others have looked at the economic impacts of a large epidemic~\cite{schoenbaum1987economic,meltzer1999economic}. However, little is known about consumer actions under a health risk, due to the lack of data encompassing such a time-concentrated outbreak~\cite{wto2020ecommerce}. This paper investigates the impact of the epidemic on online shopping behaviors and demand forecasting.

Several classical regression models exist in demand forecasting, such as the autoregressive integrated moving average (ARIMA)~\cite{contreras2003arima}. However, these models produce accurate forecasting results only when the sequence patterns are linearly correlated and stationary over time~\cite{mills1991time,omar2016hybrid}. New approaches adopt machine learning and deep learning algorithms for prediction, such as XGBoost~\cite{chen2016xgboost} and sequence-to-sequence (seq2seq)~\cite{sutskever2014sequence} models. Despite the potential, most models require mass data for training, and how they perform under a sudden disruption is yet to be investigated. This paper presents an encoder-decoder model in the prediction of near-future demands towards epidemic-related products. 


During the outbreak of COVID-19, efforts are being paid to understand COVID-19 from the perspectives of structural biology~\cite{wrapp2020cryo-em}, genetics~\cite{hoffmann2020sars}, economics~\cite{SWarren20}, policy~\cite{tian2020investigation,Maier742} and trend prediction~\cite{cohen2020scientists}. In addition to these efforts, the current study aims to provide a picture of the COVID-19 impact seen on Chinese e-commerce and leverages both the epidemic-related and behavior-related information to forecast the demand for essential goods.

\section{Data}

We use two data sources. The first is from a mobile-based shopping platform, Beidian. The platform is one of the largest in China and has a monthly user base of 3.44 million and an aggregate 187 million app downloads. It is a one-stop-shop and offers products ranging to nearly two thousand categories. We received anonymized session logs that span from January 1, 2019, to April 30, 2020. Each session information contained, for every instance, the action type (e.g., browsing, purchasing, searching), product ID, product category, and time. At the user-level, we were also given information about the cities they reside in. In the current analysis, logs originating from Hubei province were removed since the delivery of goods was prohibited during this area's lockdown.  Table~\ref{tab:basic_stats} displays the summary of data statistics. The pre-pandemic period data in (2019 and early 2020) are used to adjust for any seasonality pattern in post-pandemic period data analysis.


\begin{table}[htbp!]
    \centering
    \caption{Summary of the Beidian e-commerce dataset.}
    \setlength{\tabcolsep}{1.2mm}{
    \resizebox{0.9\linewidth}{!}{%
    \begin{tabular}{c|c|c|c|c|c}
    \toprule
        Statistics & Users & Products & Purchase & Browse & Search  \\
        \midrule
         Value & 18M & 550K & 190M & 8,285M & 1,318M \\
    \bottomrule
    \end{tabular}}}
    \label{tab:basic_stats}
\end{table}

The next data source is the daily epidemic statistics describing the newly confirmed cases within China. We combine two data sources: (1) the official reports from the Nation Health Commission of China\footnote{http://www.nhc.gov.cn/} are used for all data up to January 22, 2020, and (2) the COVID-19 dashboard data by the Center for Systems Science and Engineering at Johns Hopkins University\footnote{https://github.com/CSSEGISandData/COVID-19} are used for January 22, 2020, and onward. \looseness=-1 



\section{Modeling Market Disruptions}
\label{data_analysis}


Demands on specific health-related products such as face masks and hand sanitizers are bound to increase a health crisis, leading to a temporary surge in shopping actions. To quantify the degree of a rank change in purchase popularity of a product category $c$ over a given period of $t$, we define the Relative Popularity ($ RP $) as follows:
\begin{equation}
    RP(c,t) = \log_{10}\frac{\mathrm{ranking}(c,t_0)}{\mathrm{ranking}(c,t)},
\end{equation} 
where $t_0$ is the reference time to which a target period's popularity is compared to. We consider the first week of 2020 as a reference point and repeatedly compute the $ RP $ value every week in 2020. This metric measures the change of popularity at the product-level and address potential confounders such as the number of active users. The trajectory of this value represents disruptions on the purchase before and after COVID-19. 

\begin{figure}[h]
    \centering
    \hspace{-6mm}
    \includegraphics[width=0.47\textwidth]{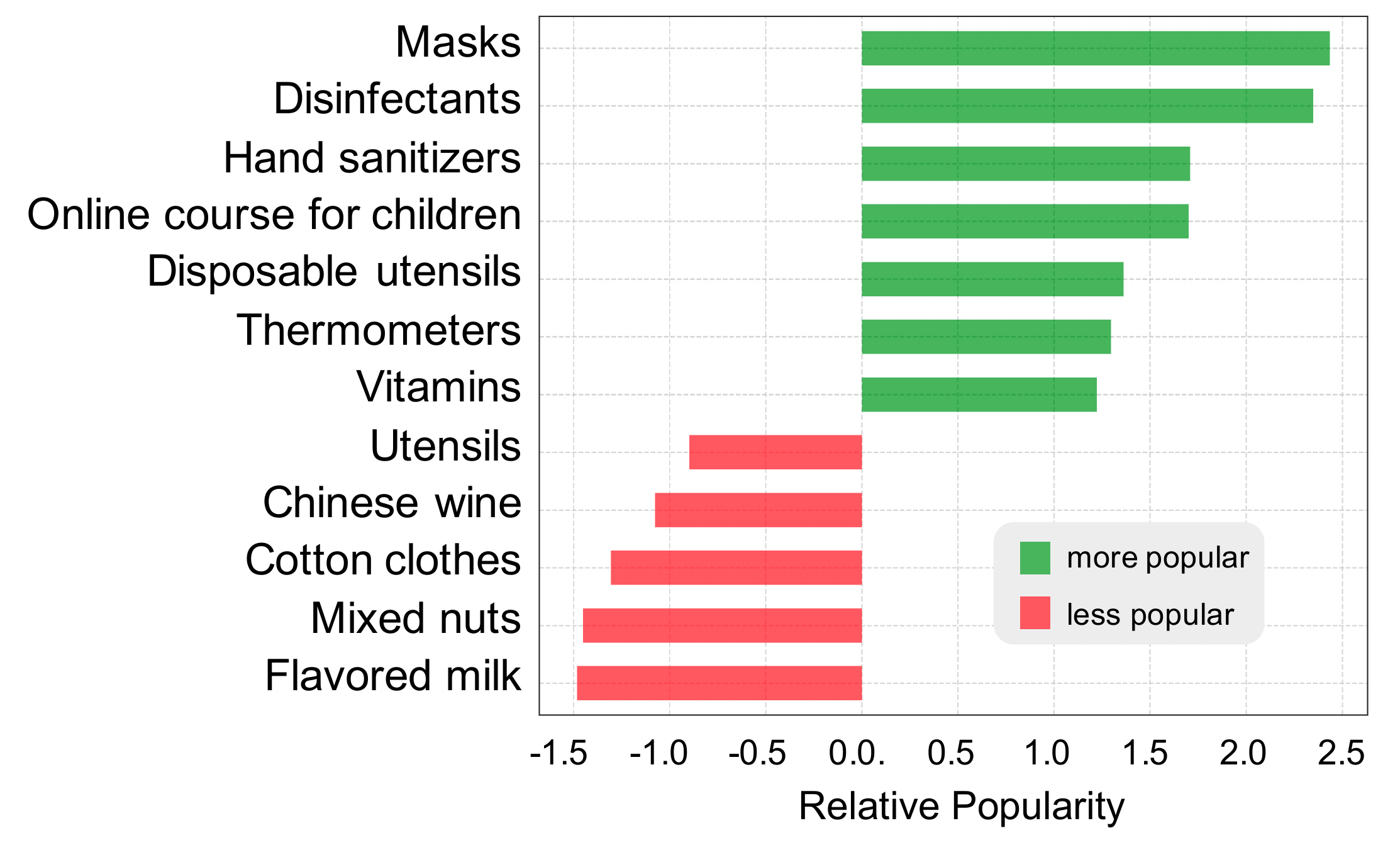}
    \vspace{-2mm}
    \caption{Products with the largest purchase rank change.
    }
    \hspace{10mm}
    \label{fig:rp_bar}
    \vspace{-5mm}
\end{figure}

\begin{figure*}[t]
\centering
\subfigure[Purchasing (week scale)]{
{\label{subfig:sales3}}
\includegraphics[width=.42\linewidth]{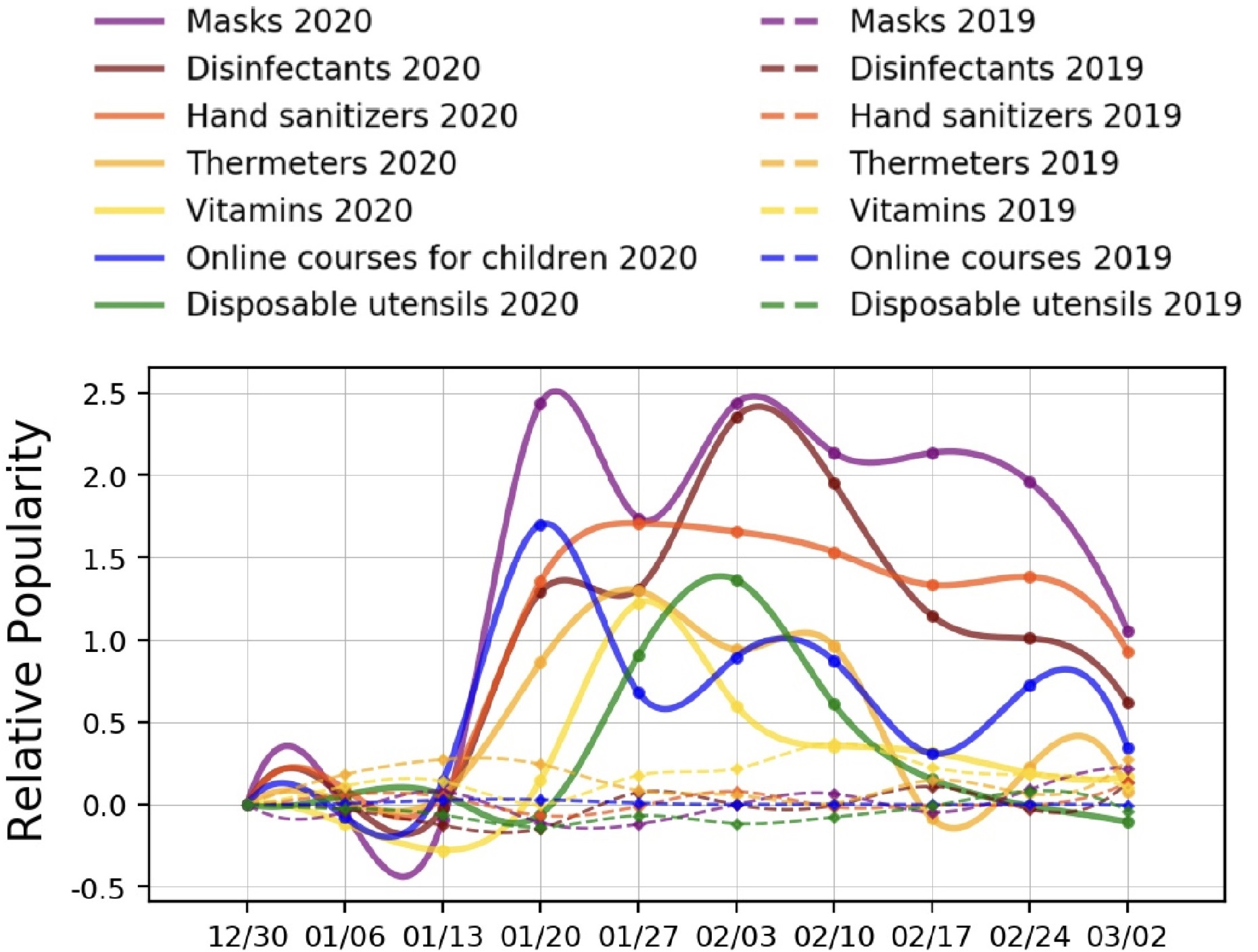}
}
\subfigure[Purchasing (week scale)]{
{\label{subfig:sales2}}
\includegraphics[width=.42\linewidth]{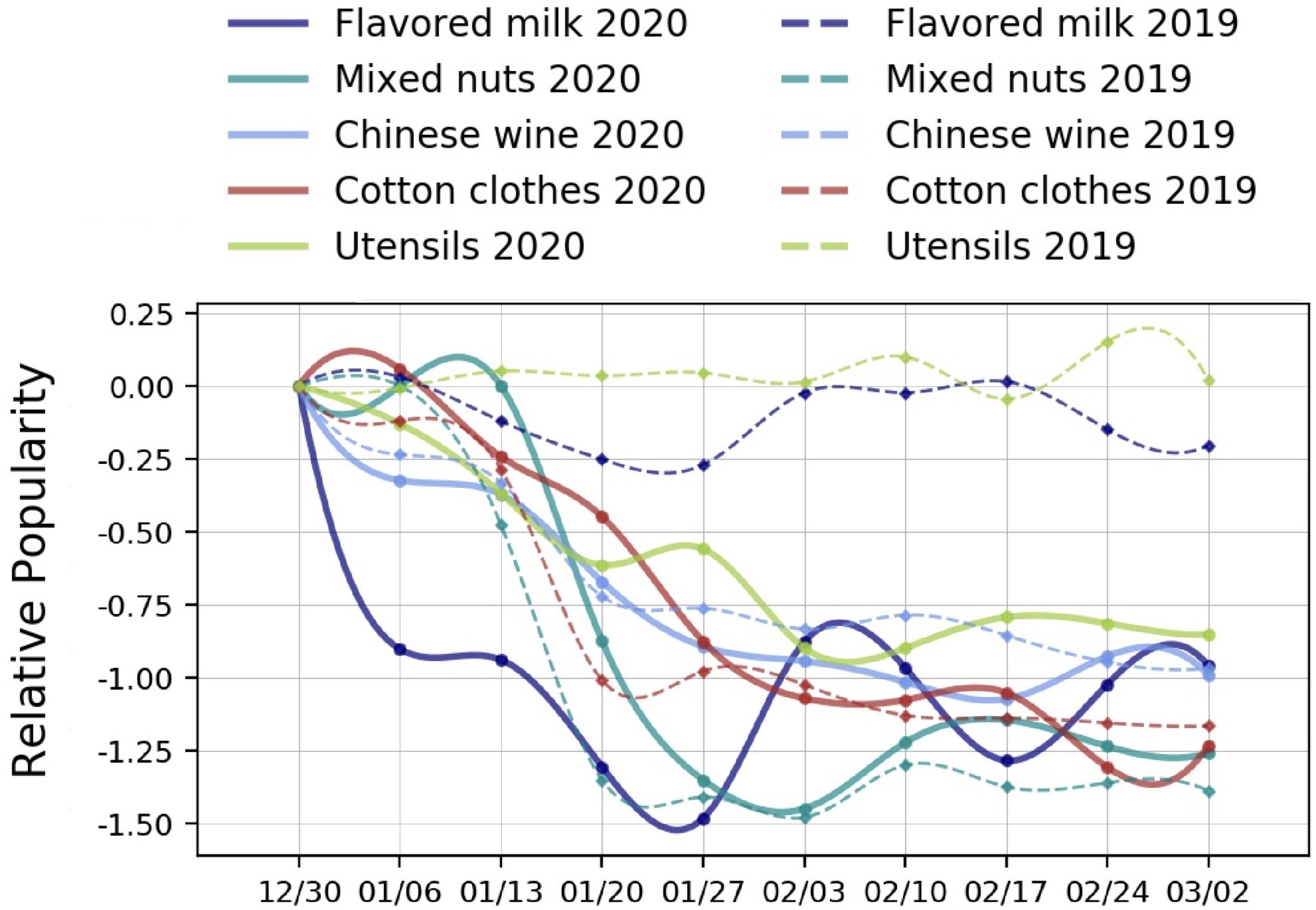}
}
\vspace*{-2mm}
\subfigure[Browsing (week scale)]{
{\label{subfig:browse3}}
\includegraphics[width=.42\linewidth]{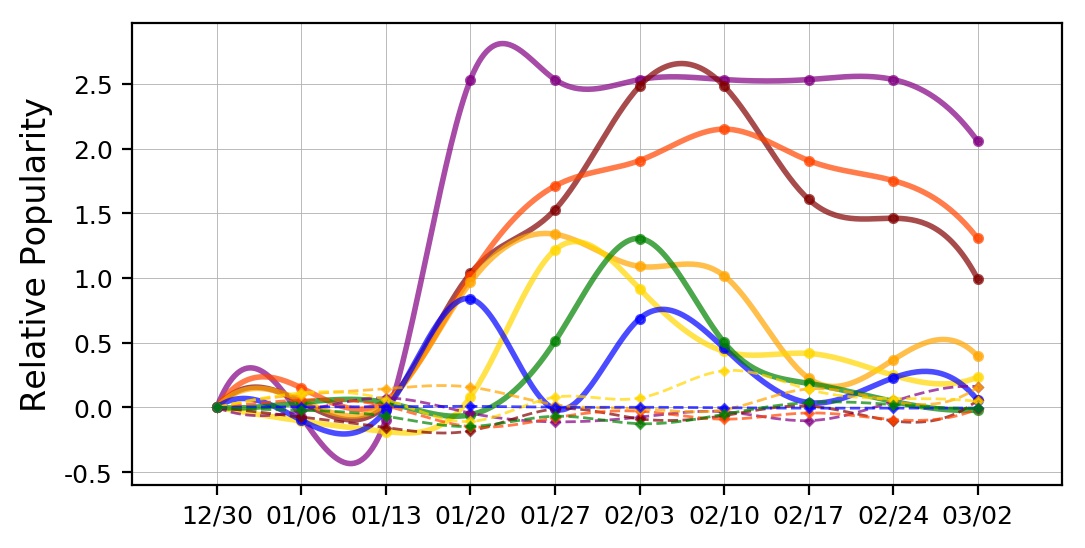}
}
\subfigure[Browsing (week scale)]{
{\label{subfig:browse2}}
\includegraphics[width=.42\linewidth]{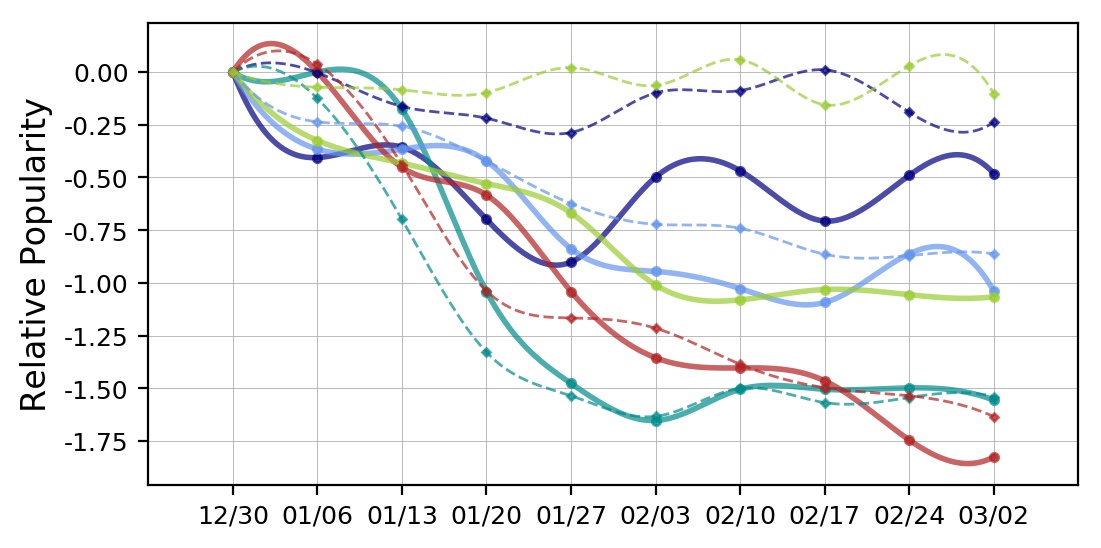}
}
\vspace*{-2mm}
\subfigure[Searching (week scale)]{
{\label{subfig:search3}}
\includegraphics[width=.42\linewidth]{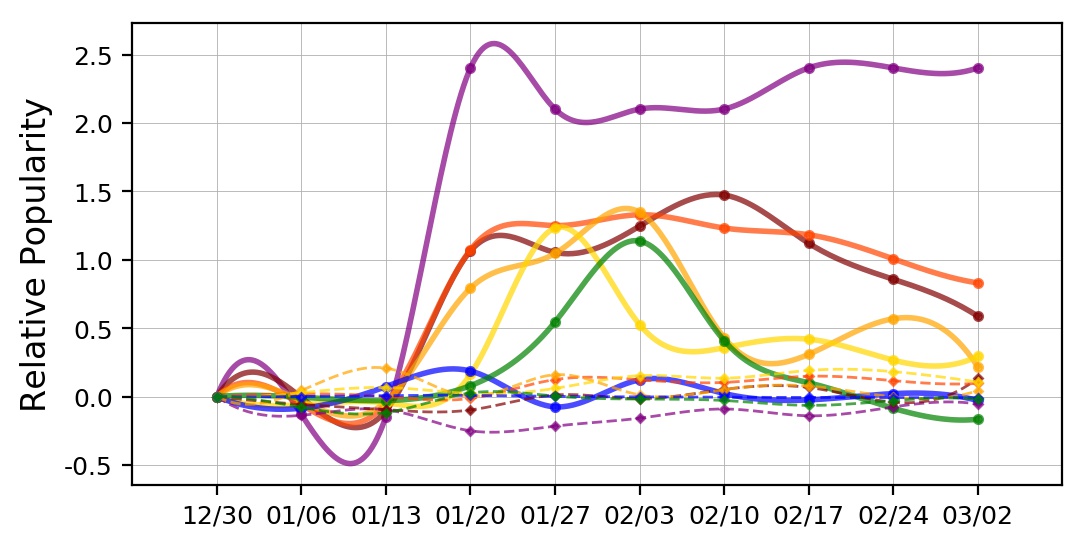}
}
\subfigure[Searching (week scale)]{
{\label{subfig:search2}}
\includegraphics[width=.42\linewidth]{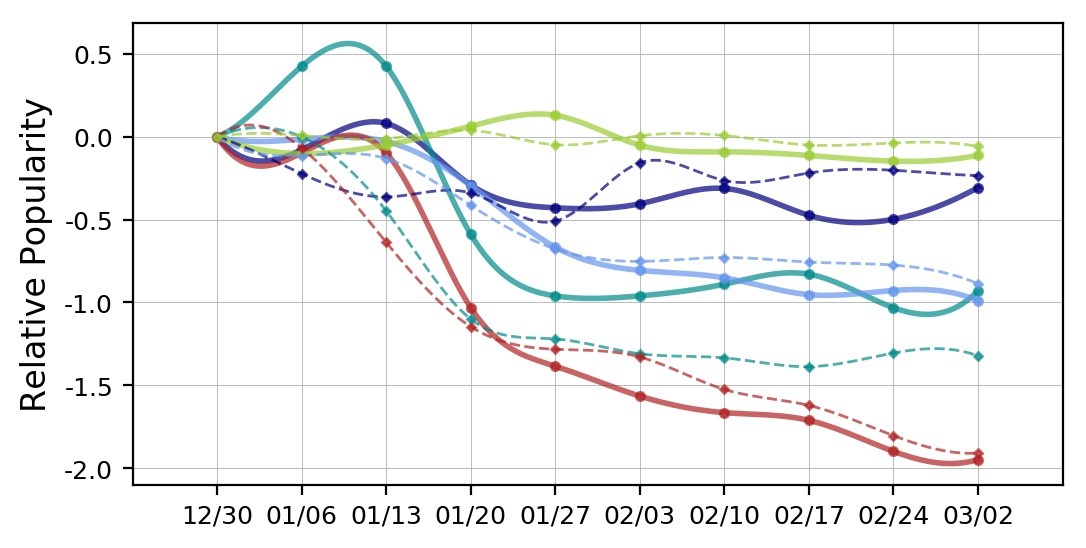}
}
\caption{
Weekly popularity dynamics during the COVID-19 period. Most products in panels (a), (c) are increasing in their relative popularity ranking, whereas most products in panels (b), (d) show decreasing popularity ranks.
}
\label{fig:weekly_ranking}
\vspace*{-3mm}
\end{figure*}

\subsection{Popularity change by products}\label{behavior_analysis}

\subsubsection{The most affected products.}
Figure~\ref{fig:rp_bar} presents the signature categories that mark the highest and the lowest $RP$ values. The top seven items, from masks to vitamins, are within the top-20 to increase purchase rank change; their sales have surged compared to the final week of 2019. In contrast, the bottom five items from utensils to flavored milk, within the bottom-20, mark the largest decrease in purchase rank change; their sales decrease the largest during the pandemic. 
%


\begin{figure*}[t]
\centering
\hspace*{-10mm}
\subfigure[Masks (week scale)]{
{\label{subfig:masks}}
\includegraphics[width=.4\linewidth]{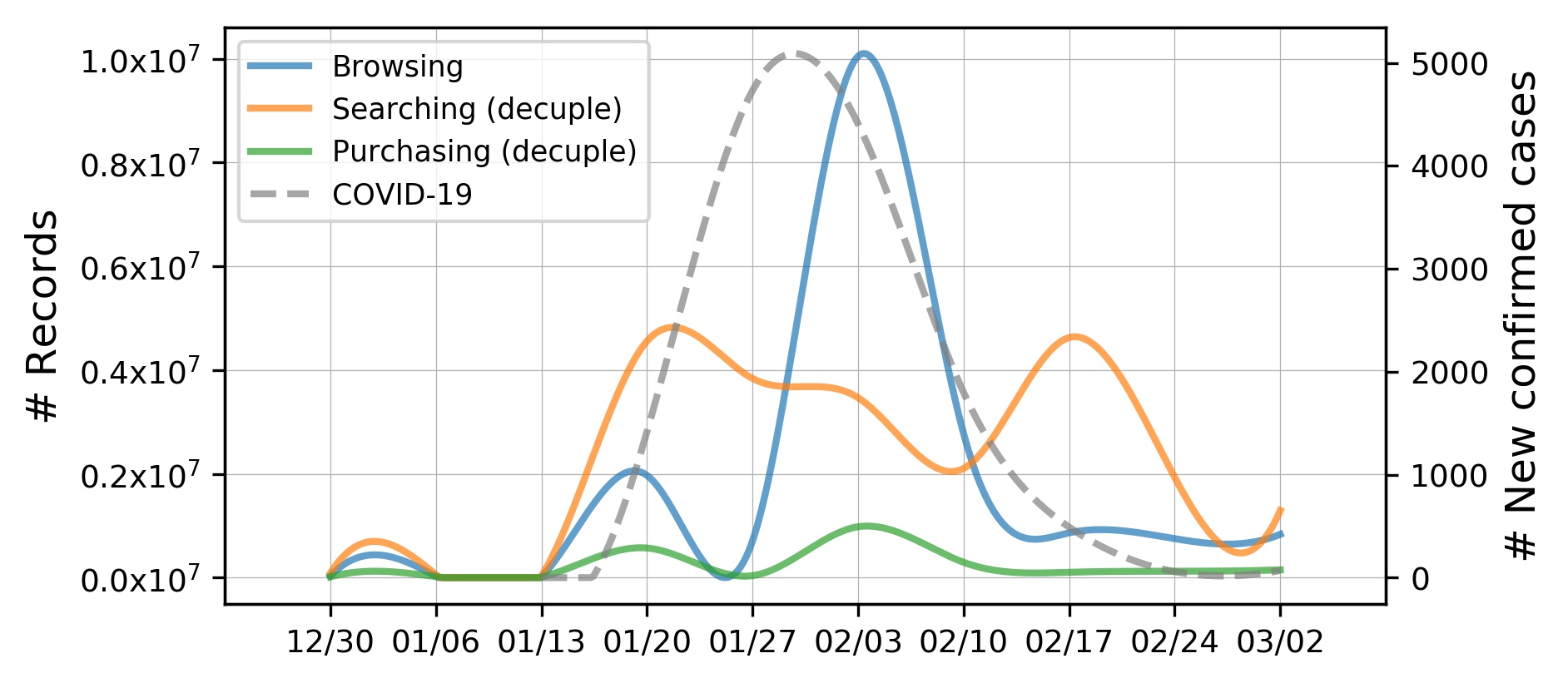}
}
\subfigure[Disinfectants (week scale)]{
{\label{subfig:disinfectant}}
\includegraphics[width=.4\linewidth]{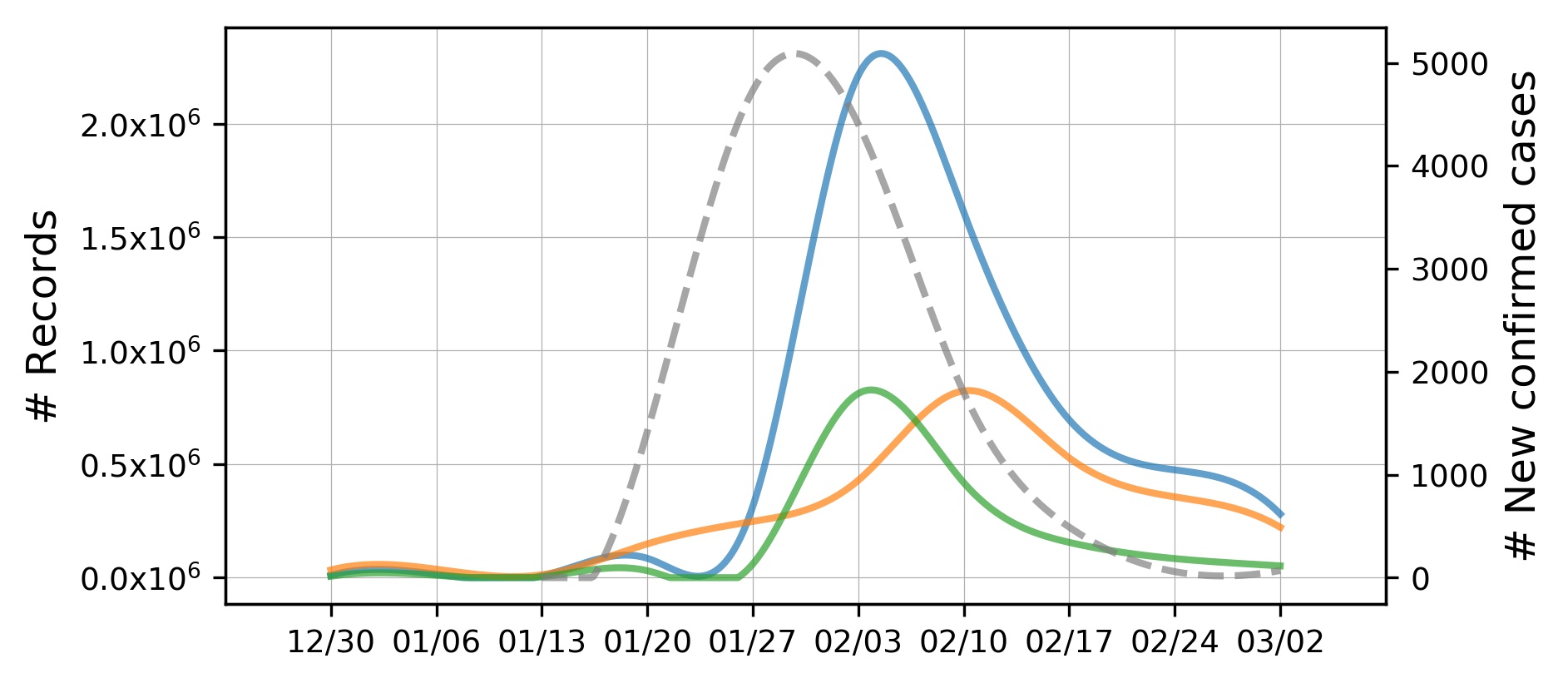}
}
\hspace*{-10mm}
\hspace*{-10mm}
\subfigure[Hand sanitizers (week scale)]{
{\label{subfig:hand_sanitizers}}
\includegraphics[width=.4\linewidth]{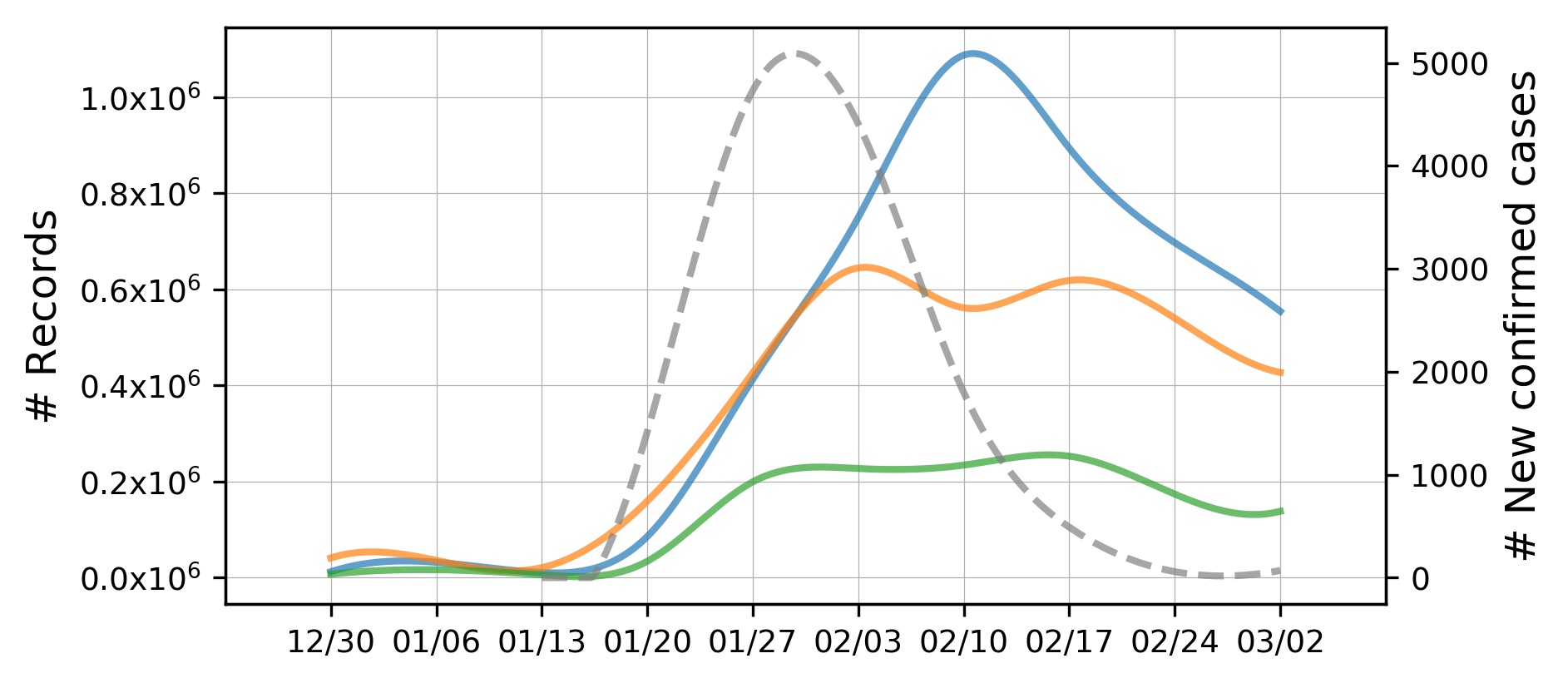}
}
\subfigure[Mixed nuts (week scale)]{
{\label{subfig:milk}}
\includegraphics[width=.4\linewidth]{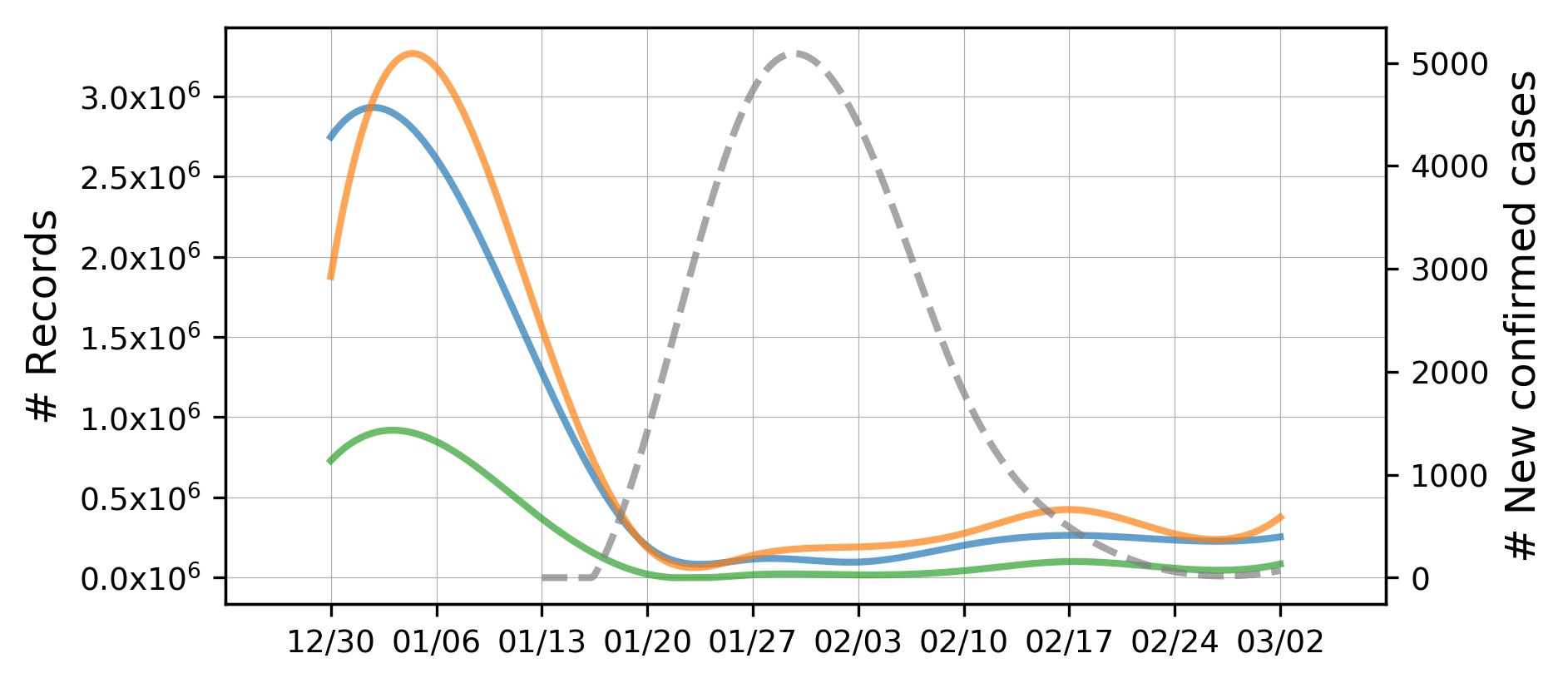}
}
\hspace*{-10mm}
\vspace*{-2mm}
\caption{Weekly records of browsing, searching and purchasing on different kinds of products. Those of searching and purchasing are multiplied by ten (i.e., $1 \times 10^5$ in these figures actually represents $1 \times 10^4$ of searching and purchasing).
}\label{fig:action_comp}
\vspace*{-4mm}
\end{figure*}

The top list includes epidemic-related products such as face masks, disinfectants, hand sanitizers, and thermometers. The list also includes online course programs for children, disposable utensils, and vitamins, which could be assumed relevant during the pandemic given homeschooling, hygiene, and immunity have become either mandatory or important. We see (non-disposable) utensils, Chinese wine, cotton clothes, mixed nuts, and flavored milk moving on to the bottom list. These items are mostly within the top 100 before the pandemic. However, their purchase ranks drop by more than ten times during the pandemic. We list the full list of top-20 and bottom-20 product categories in the Appendix. \looseness=-1

\subsubsection{Weekly fluctuations.}
Figure~\ref{fig:weekly_ranking} shows the week-by-week $RP$ values of the 12 prominent goods along with their fitted lines. Here $t_0$ is again set to the first week of 2020. Product rank is shown for all three shopping actions: browsing, searching, and purchasing. The figure also shows the popularity trajectory of the same items in the year of 2019. We shift the timeline for 2019 and sync the new year's holiday week to appear as the fourth data point, to discount seasonal effect. Note that the studied popularity measure $ RP $ is stable and applicable to all popularity levels since it looks at the relative rank changes in the logarithmic scale. 

Products in panels~(a) and (c) such as face masks, disinfectants, hand sanitizers, and thermometers show a surge in all shopping actions from the week of January 20th.\footnote{The surge appears on the fourth data points; the fitted lines are for visual aid and do not represent a gradual increase in popularity.} This is when the lockdown of Hubei province was enforced. Face masks, in particular, continue to remain the top-ranked item throughout the pandemic period. The rank difference is substantially high and above 2.0 for the search action. Note that the $RP$ value of 2.0 indicates several hundreds of rank increases. The dashed lines, representing the identical item's rank changes in 2019, do not show any notable increase. This confirms that the surges are not seasonal but unique to 2020 (i.e., epidemic-related).

The sudden rank change for children's online courses is also noticeable during the pandemic's first week. The search action of this product, nonetheless, is not as high as other top products. We also pay attention to the products like thermometers, vitamins, and disposable utensils whose peaks in demand come a week or two after the disease outbreak. The times at which each product shows the highest rank change in demand could be used to understand what popular health practices households adopt during the pandemic and how much they are concerned about the disease.


In contrast, products in panels~(b) and (d) show decreased ranks in all shopping actions. Some changes, however, are seasonal and can also be observed in 2019. For instance, the rank order of cotton clothes and Chinese wine gradually decreases in 2019 and 2020, indicating a seasonal effect (e.g., warmer weather, wine demands decreasing after the new year's celebration). However, products like flavored milk and utensils show decreased popularity only in 2020, indicating households consume certain items less during the pandemic. It is interesting to contrast the decreasing demand for utensils against disposable utensils that shows a rise in demand during the pandemic. This contrast likely arises due to increased efforts for hygiene.

\begin{figure*}[htbp!]
\centering
\hspace{-3mm}
\subfigure[Masks]{
{\label{subfig:mask_purchase}}
\includegraphics[width=.225\linewidth]{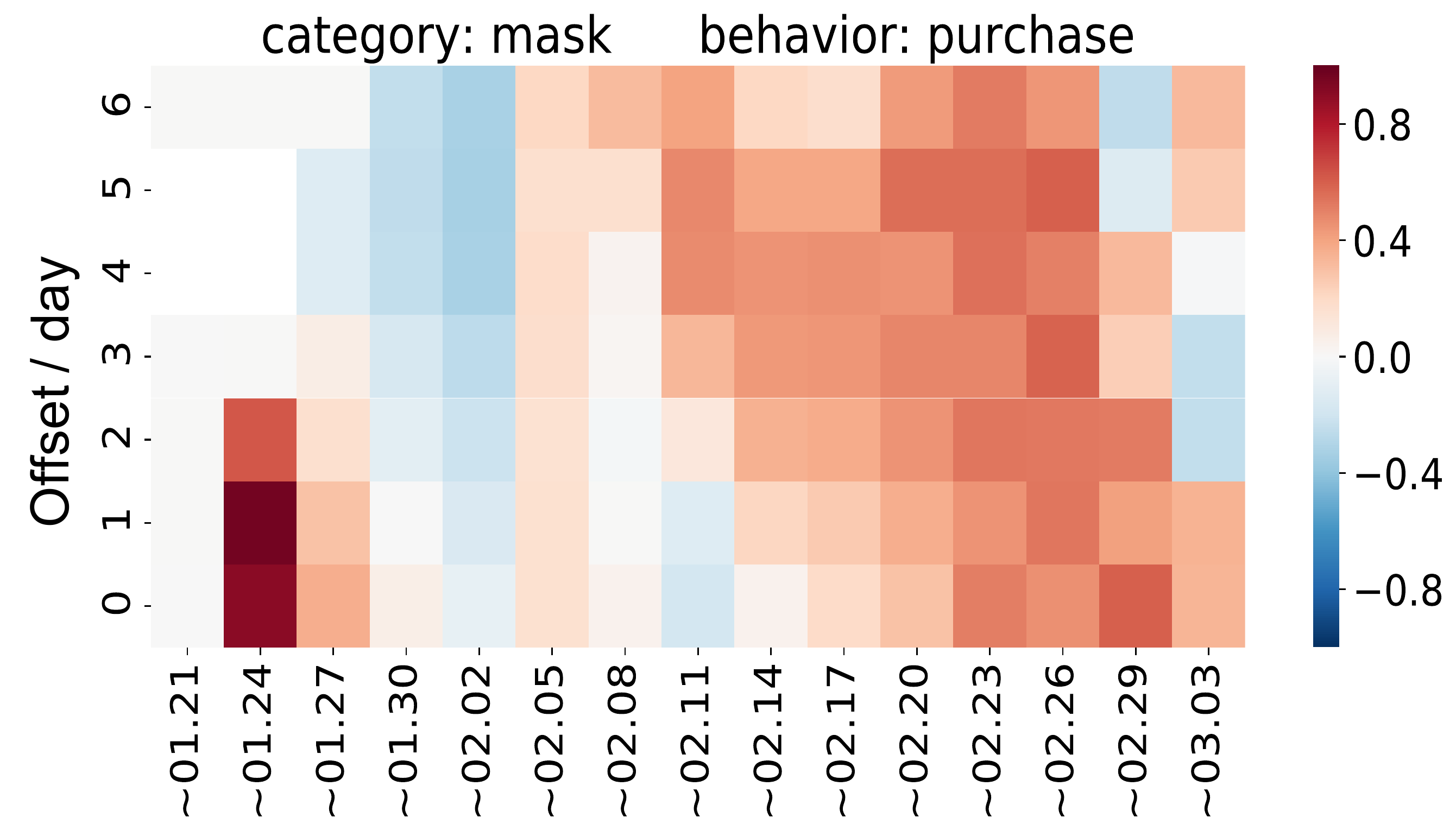}
}
\hspace{-3mm}
\subfigure[Disinfectants]{
{\label{subfig:disinfectant_purchase}}
\includegraphics[width=.24\linewidth]{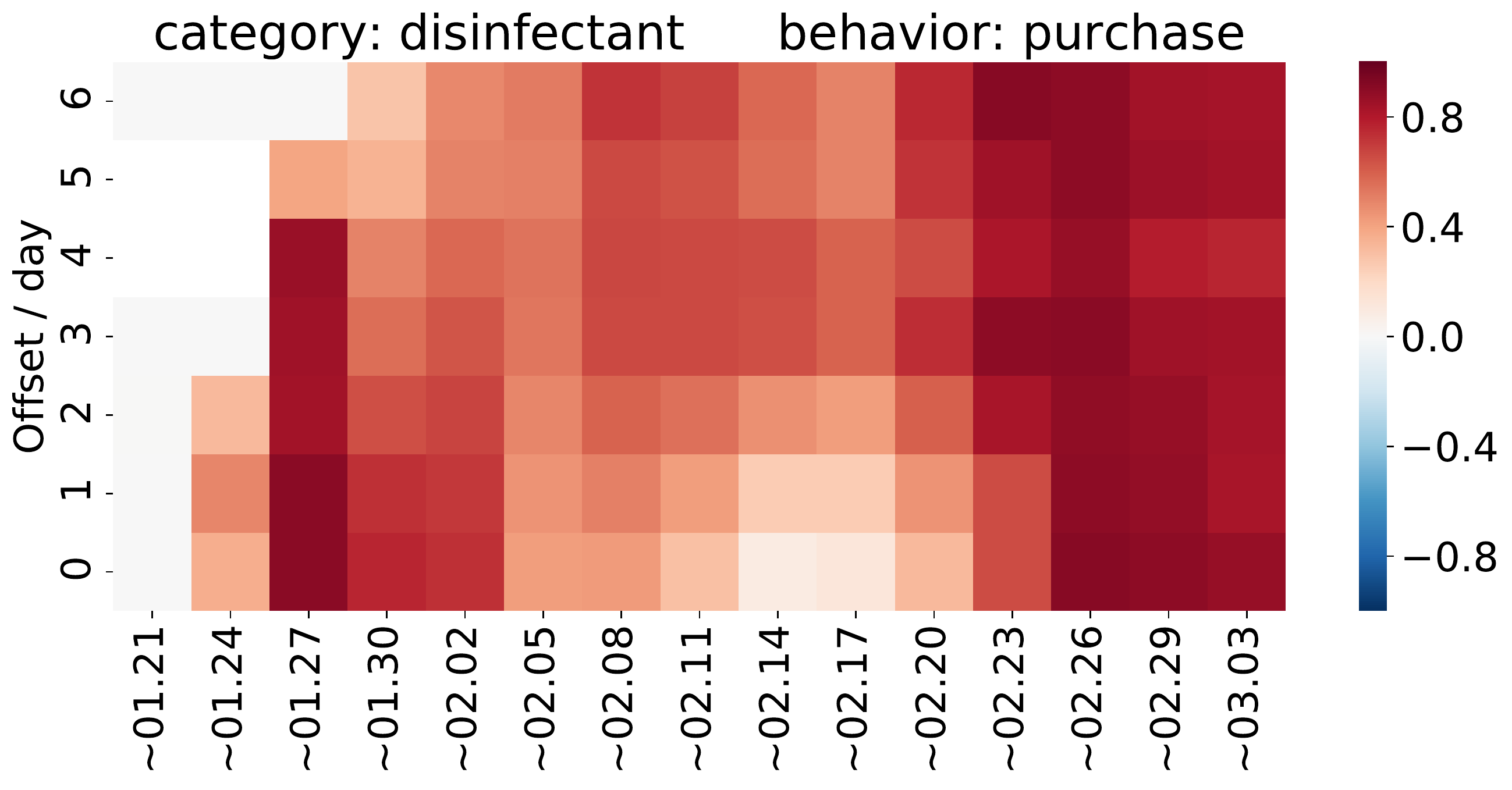}
}
\hspace{-3mm}
\subfigure[Hand sanitizers]{
{\label{subfig:hand_sanitizer}}
\includegraphics[width=.24\linewidth]{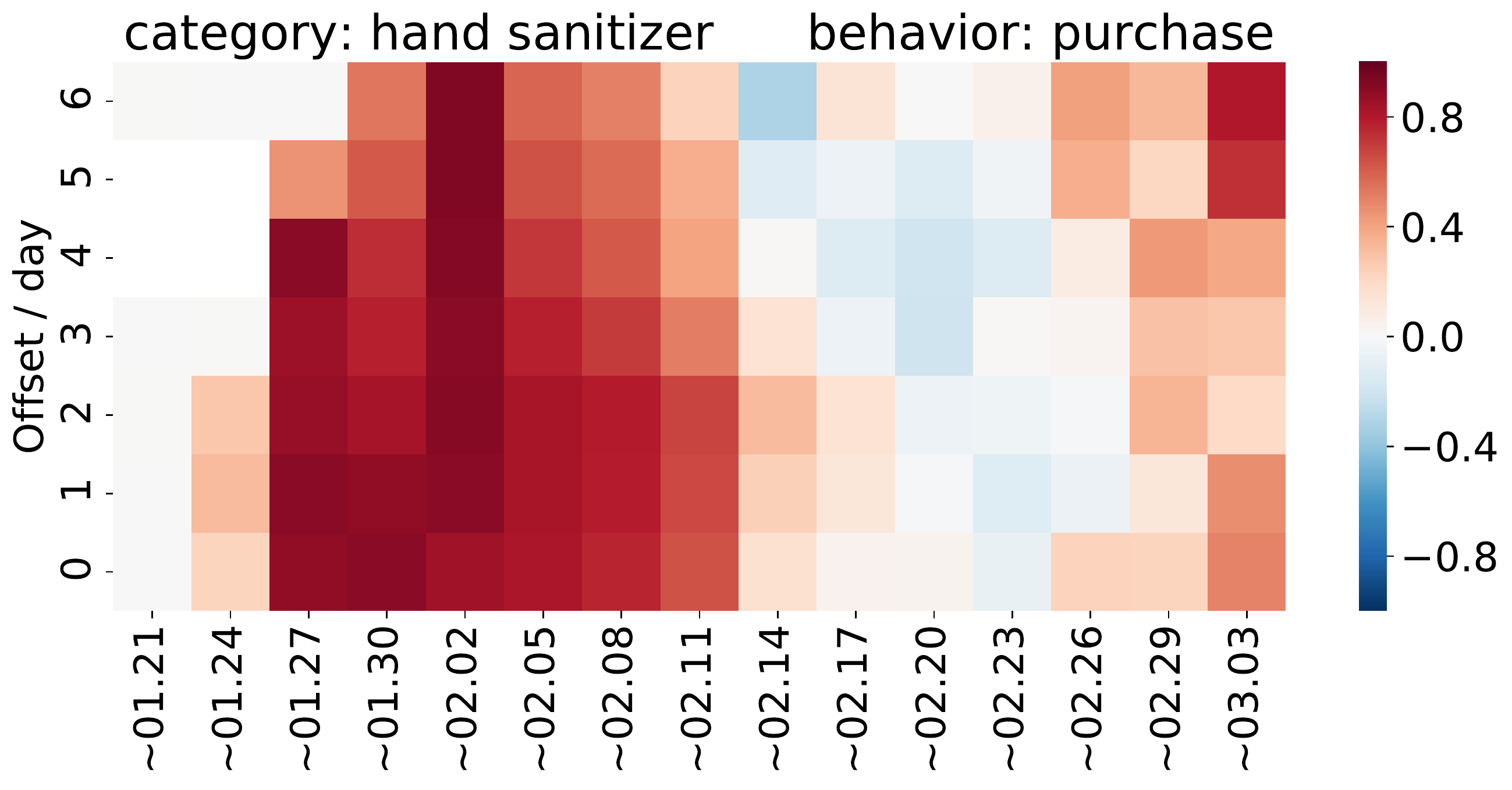}
}
\hspace{-3mm}
\subfigure[Vitamins]{
{\label{subfig:vitamin}}
\includegraphics[width=.24\linewidth]{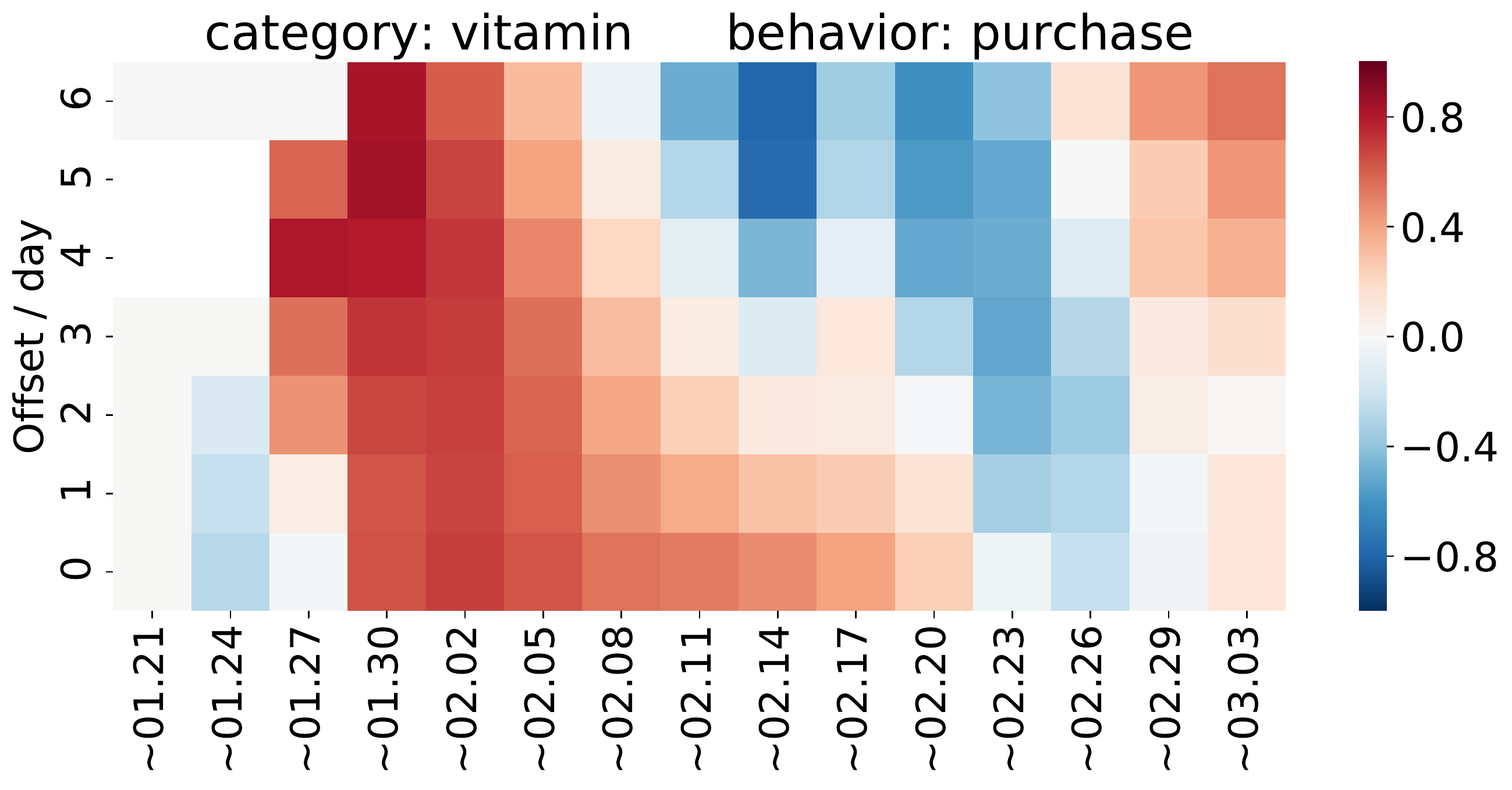}
}
\vspace*{-1mm}
\caption{Time-lagged cross-correlation results of different products with a rolling time window. The color denotes the Pearson correlation coefficient between the two time series of corresponding behavior and newly confirmed cases. 
}\label{fig:categories}
\vspace*{-2mm}
\end{figure*}

\subsubsection{Session counts by action type.}
To examine the dynamics across the shopping actions, Figure~\ref{fig:action_comp} compares the session counts for four representative products. The data shows that shoppers engage the most frequently in browsing actions, up to ten times more than the searching and purchasing actions. Also shown in the figure is the number of weekly confirmed epidemic statistics in China. 

For mixed nuts, a non-epidemic product, the relative session counts across the three action types remain similar over time. In contrast, for the other three epidemic-related goods, searching and purchasing action rates may differ because products may be sold out. Interestingly, browsing and purchasing actions show similar fluctuations. These distinctive patterns are the most obvious for face masks in Figure~\ref{subfig:masks}, where the searching action demonstrates that it continues to be in high demand. Plots of disinfectants show a similar trend. The plot of hand sanitizers, however, show less difference across the three shopping actions. This may be because hand sanitizers were accessible more widely than the other two items.  



\subsection{Time-lagged analysis
}\label{TLCC}

The temporal analysis so far has revealed the dependency of online shopping actions to COVID-19; demand for some products is immediate upon a health risk, but other products stagger in their rank change. This subsection examines the time difference in detail via computing the lagged correlation between the two sequences. We compute the time-lagged cross-correlation (TLCC)~\cite{shen2015analysis}. 
The method shifts two sequences relatively in time (time-lagged) and calculates the correlation between them (cross-correlation). Therefore, it can analyze non-stationary time series and quantify how the three shopping actions appear upon a pandemic. 
We use a rolling window to analyze different periods and set each window to be a size of 21 days, and the lagged time is set from 0 to 6 days between the two series. Figure~\ref{fig:categories} presents the results of the time-lagged cross-correlation (TLCC) for selected products based on the purchase log. The y-axis represents the time-difference offset denoting the number of days that behavioral response falls behind the epidemic spreading, and the color represents the Pearson correlation coefficient from negative (blue) to positive (red).

The correlations are not uniform across products. Figure~\ref{subfig:mask_purchase} demonstrates that the strong correlation between mask purchases and epidemic development does not last long due to falling mask supplies. This trend is shown by dark red blocks on Jan 14, followed by blue shaded regions in the figure. Figure~\ref{subfig:disinfectant_purchase} in contrast, shows that disinfectant purchases closely follow the epidemic development throughout the whole period, resulting in high correlation (i.e., red shaded areas). We can deduce that there is no shortage of disinfectant supplies on the platform. 

On the other hand, the demands of hand sanitizers(Figure~\ref{subfig:hand_sanitizer}) and vitamins(Figure~\ref{subfig:vitamin}) remain high until the end of February. For example, as shown in Figure~\ref{subfig:hand_sanitizers} and \ref{subfig:hand_sanitizer}, sales of hand sanitizer skyrocket by following the epidemic development at the early stages of COVID-19. It shows a week's lagged response to epidemic development. However, the lagged effect decreases over time, implying that shopping demands become neutral to epidemic development. Furthermore, a positive relationship becomes weak and turns to a negative. This can be interpreted as hand sanitizers are still high in demand, even after the number of new confirmed cases decreased. Toward the end of the time frames, the lagged positive correlation appears again, which means that consumers have stocked enough hand sanitizers, thus no longer purchase them.

\begin{figure}[htbp!]
\centering
\subfigure[Searching on masks]{
{\label{subfig:mask_search}}
\hspace{-6mm}
\includegraphics[width=.49\linewidth]{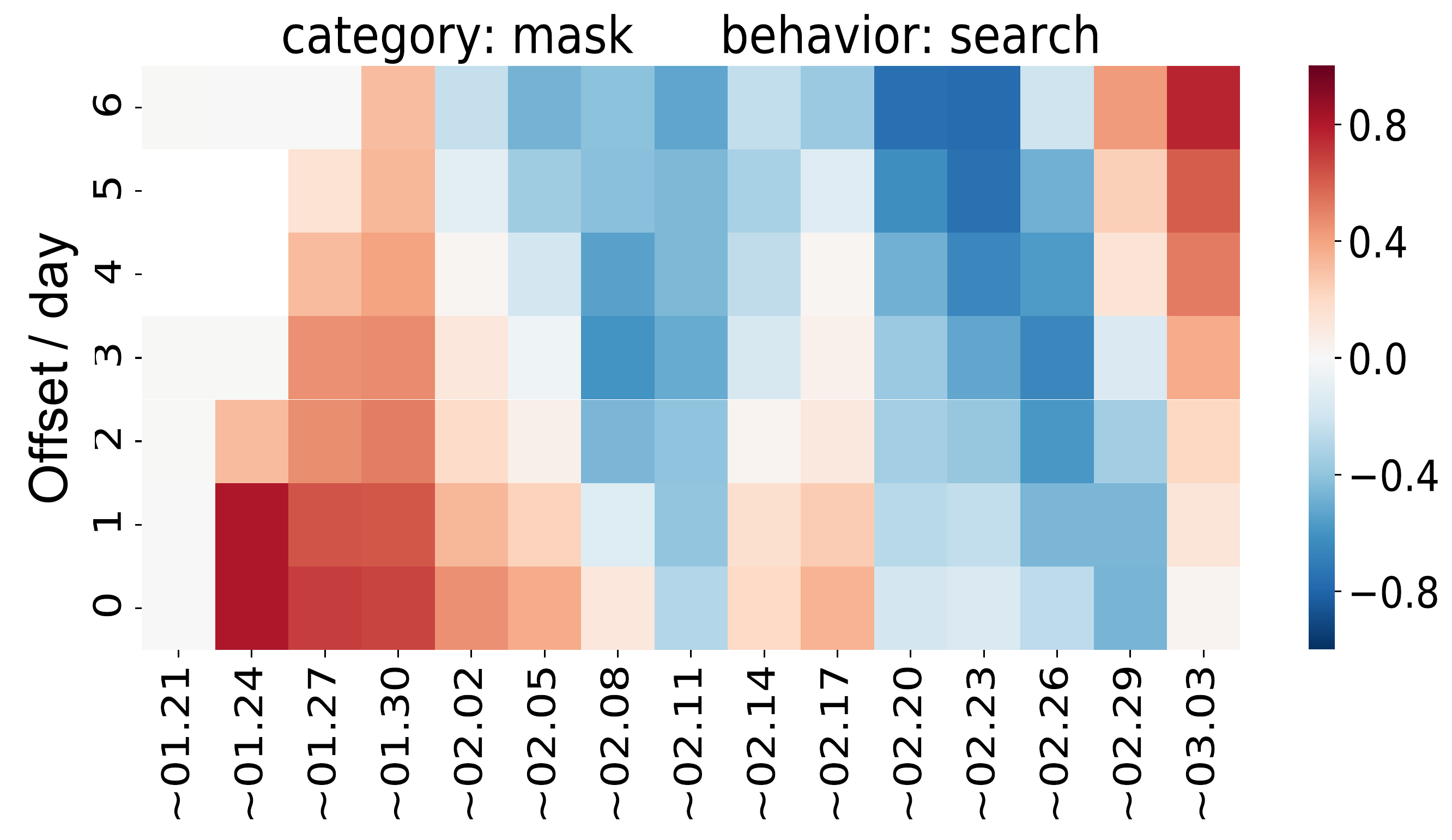}
}
\hspace{-1mm}
\subfigure[Purchasing on masks]{
{\label{subfig:mask_sales}}
\hspace{-6mm}
\includegraphics[width=.49\linewidth]{figures/TLCC/mask_sales.pdf}
}
\vspace{-1mm}
\caption{Even when the epidemic subsided, consumer interest in searching masks remained. For preparing future epidemic, demand forecasting and inventory planning are vital to avoid supply shortage.}\label{fig:behaviors}
\end{figure}

Our data shows that correlations are found across shopping actions. Mask is a representative product that faced supply shortage during the epidemics in China. The searching actions in Figure~\ref{subfig:mask_search} show a negative correlation in the later period of epidemics, indicating that people are still searching for masks although the COVID-19 confirmed cases is decreasing, identified as blue shaded regions. Positive correlation appearing 
on the later part of Figure~\ref{subfig:mask_sales} can be understood in the same context that only when the number of confirmed patients decreased, the number of sales and the number of confirmed cases show positive correlation.

In summary, our analysis shows that dynamic correlations exist between online shopping behaviors and epidemic development. We find that behaviors respond to the epidemic in a lagged manner, but the correlation can be reversed when there is a shortage or continued caution. These patterns and observations inspire us to design an accurate and explainable predictor for forecasting consumer demand on key product categories. \looseness=-1
\section{Demand Forecasting and Evaluations}\label{method}


The analysis so far has demonstrated how COVID-19 impacted product popularity and behavioral patterns. The distinctive patterns of the purchasing and searching for critical items like facial masks suggest that the pandemic disrupts the supply of essential goods and leads to an imbalance of demand and supply. The analysis also confirmed a significant correlation between shopping actions and epidemic sequences. Together, these findings suggest that many products' purchase intent is directly affected by the epidemic development during a health crisis. 
\begin{figure}[bhtp!]
    \centering
    \includegraphics[width=0.45\textwidth]{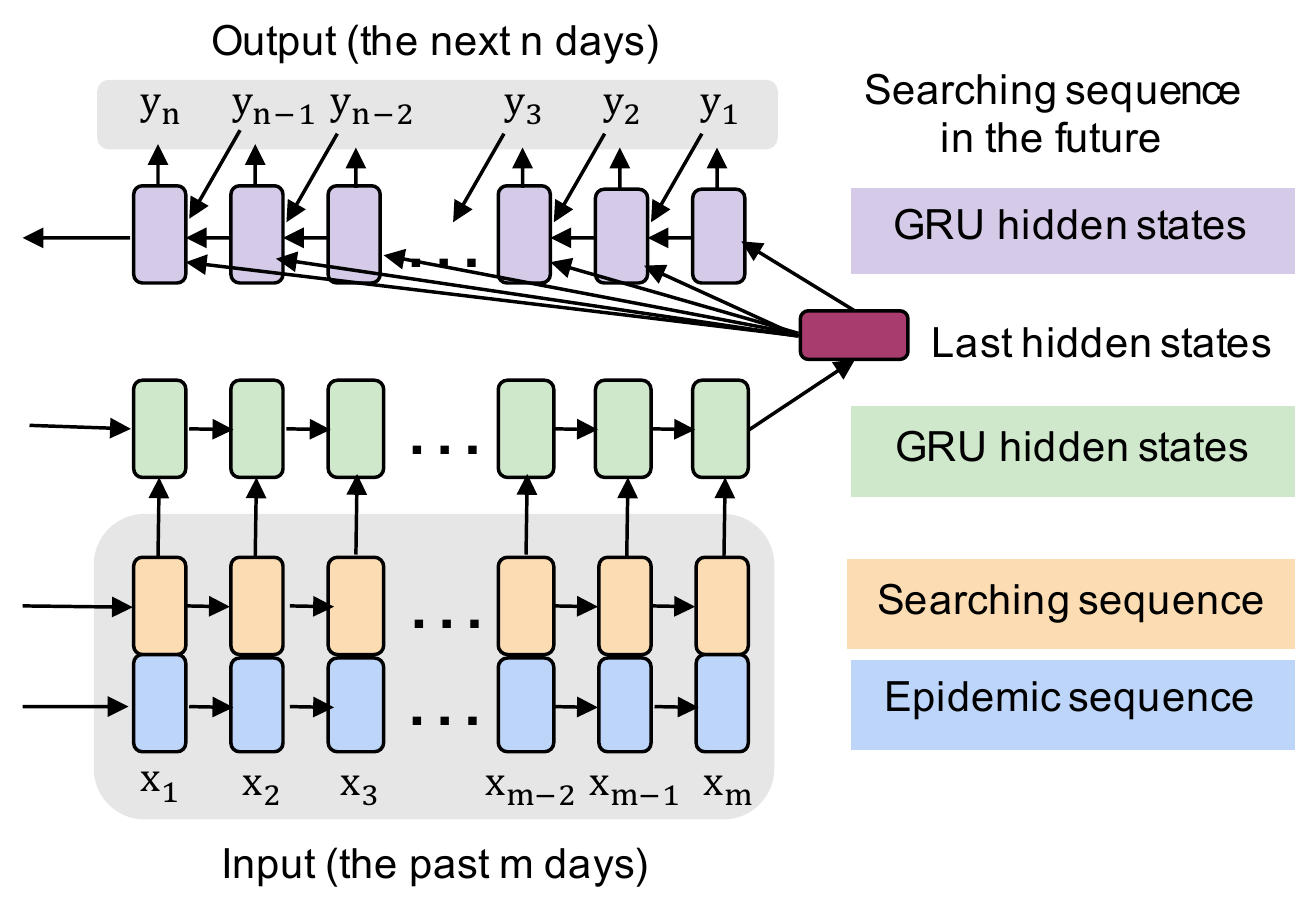}
    \caption{The overall architecture of our proposed model EnCod, which takes in the concatenation of epidemic and searching sequences in the past $m$ days and outputs the daily number of searches in the next $n$ days.}
    \label{fig:model_all}
\end{figure}
%

\begin{table*}[htpb!]
    \centering
    \caption{The NRMSE performance of our method compared to baselines. The results represent the prediction performance in China except for Hubei province.} 
    \resizebox{\textwidth}{!}{
    \begin{tabular}{c | c c c c c | c c c c c }
    \toprule
     & \multicolumn{5}{c|}{\textbf{Products with increased purchase ranks (COVID-19 related)}}  & \multicolumn{5}{c}{\textbf{Products with decreased purchase ranks (COVID-19 irrelevant)}}  \\
    \cmidrule{2-6}\cmidrule{7-11}

    \multirow{2}{*}{\textbf{Model}} &   &   & \textbf{Hand} &   & \textbf{Thermo-}  & \textbf{Flavored} & \textbf{Daily} &  & \textbf{Mixed} & \\
    
    ~ & \textbf{Masks} & \textbf{Disinfectants} & \textbf{sanitizers} & \textbf{Vitamins} & \textbf{meters} & \textbf{milk} & \textbf{necessities} & \textbf{Vegetables} & \textbf{nuts} & \textbf{Decorations}\\

    \midrule
    
    AR & 0.657 & 0.366 & 0.226 & 0.287 & 0.248 & 0.281 & 0.287 & 0.387 & 0.269 & 0.469 \\
    ARIMA & 0.461 & 0.633 & 0.865 & 0.258 & 0.232 & 0.292 & 0.298 & 0.405 & 0.267 & 0.447 \\
    Seq2seq & 0.438 & 0.323 & 0.197 & 0.216 & 0.149 & 0.283 & 0.305 & 0.330 & 0.253 & 0.286 \\
    XGBoost & 0.718 & 0.578 & 0.597 & 0.492 & 0.309 & 0.283 & 0.301 & 0.327 & 0.294 & 0.463 \\
    XGBoost-C & 0.468 & 0.512 & 0.289 & 0.307 & 0.293 & 0.281 & 0.330 & 0.326 & 0.295 & 0.547 \\
    
    \midrule
    
    \textbf{EnCod} & \textbf{0.307} & \textbf{0.212} & \textbf{0.156} & \textbf{0.201} & \textbf{0.126} & 0.285 & 
    0.279 & 0.331 & 0.252 & 0.284 \\

    \bottomrule
    \end{tabular}
    }
    \label{tbl:performance}
\end{table*}


Based on the insights above, we present a Gated Recurrent Unit (GRU)-based encoder-decoder model named \textbf{EnCod} that leverages the epidemic information along with historical shopping behaviors to predict the demand for critical goods. We use the data of daily confirmed cases and searching behaviors in the past two weeks to predict the number of searches in the following $n$ days. Figure~\ref{fig:model_all} shows the detail of the prediction model. The EnCod model is based on the GRU network~\cite{bahdanau2014neural}. The model takes in the concatenation of sequences of daily confirmed cases and searches as input. The encoder module extracts the historical sequence features and outputs the last hidden states that serve as the decoder's input. Then the decoder model generates the prediction results in the future.

\subsection{Performance evaluations}

We compare our model with baselines of classical time series forecasting algorithms, including autoregressive (\textbf{AR}) model~\cite{mills1991time},  \textbf{ARIMA} model~\cite{contreras2003arima}, and a deep learning algorithm like \textbf{Seq2Seq} and \textbf{XGBoost}~\cite{chen2016xgboost}. These methods only use in historical shopping behaviors and predict its future. For a fair comparison, we also use a variant of the XGBoost model that utilizes both the historical shopping history and epidemic statistics (which we call \textbf{XGBoost-C}).

We evaluate the prediction performance with the Normalized Root Mean Square Error (NRMSE)~\cite{rocha2007evolution}. The experimental period is from January 1 to March 31, 2020, covering the leading COVID-19 epidemic development in China. We use historical searching logs from the Beidian platform, and the daily confirmed cases over the past two weeks to predict consumer demand (i.e., product searches) in the immediate week. We split the data into the training and testing sets with a ratio of 3:1 according to time. We train the model parameters with Adam optimizer regularized by an early stop and set the mini-batch size 10. The learning rate is initialized as 1e-2, which is gradually reduced by 0.1.

Table~\ref{tbl:performance} contains the prediction performance of EnCod along with other baselines. We choose ten categories of two groups, including both the COVID-19 related and COVID-19 unrelated product categories. Products with the highest relative popularity ($RP$) such as face masks appear in the COVID-19 related products, and those with low $RP$ value appear as the irrelevant group.\footnote{We show the full results for all of the major product categories in the Appendix.}


The COVID-19 related group results show that adding the epidemic statistics contributes to a substantial demand forecasting performance. This was consistent for both XGBoost and encoder-decoder models. (Seq2Seq is a variant of our model without the epidemic information.) Compared to the best performing baseline, EnCod could reduce the NRMSE value by an additional $6.9$\% to $34.3$\% in the prediction tasks for the COVID-19 related products. However, having the epidemic information as in XGBoost-C is not necessarily the second-best performing alternative. Sometimes it was the AR, ARIMA, or the Seq2Seq model that produced a good alternative, excluding EnCod. 

Next, in the comparison of COVID-19 irrelevant products, EnCod no longer produces the best results in all prediction tasks. EnCod is only marginally better, and the XGBoost-C model produces the best result for flavored milk and vegetables. This means that obtaining additional epidemic statistics is no longer helpful for items with low $ RP $ value. Among the items, we show the results for daily necessities (e.g., toilet paper, storage bags, kitchen supplies) and home decorations. All other results can be found in the Appendix. 

The comparison results between the two groups validate that our method can capture the relationship between epidemic development and the demand change of essential goods. Moreover, it verifies the usability of the metrics $ RP $, which is defined to characterize the market. The prediction results indicate that the metrics $ RP $ can distinguish the product's relevance to an epidemic.


\subsection{Regional and long-term forecasting}\label{regional}

\begin{table*}[htbp!]
    \centering
    \caption{The NRMSE performance of baseline methods and our model measured in nine representative provinces in China. Provinces nearby Hubei are marked with the $^*$ sign. }
    \resizebox{0.8\textwidth}{!}{
    \begin{tabular}{c|c c c c c c c c c}
    \toprule
    \multirow{2}{*}{\textbf{Model}} & \multicolumn{9}{c}{\textbf{Provinces}} \\
    \cmidrule{2-10}
    
    ~ & \textbf{Beijing} & \textbf{Shanxi} & \textbf{Jilin}  & \textbf{Zhejiang$^*$} & \textbf{Shanghai} & \textbf{Sichuan$^*$} & \textbf{Guangdong} & \textbf{Hunan$^*$} & \textbf{Henan$^*$}\\
    
    \midrule
    
    AR  & 0.521 & 0.990 & 0.686 & 0.454 & 0.597 & 0.432 & 0.990 & 0.398 & 0.563 \\
    ARIMA & 0.442 & 0.491 & 0.536 & 0.346 & 0.439 & 0.340 & 0.589 & 0.350 & 0.345 \\
    Seq2seq & 0.338 & 0.403 & 0.579 & 0.267 & 0.353 & 0.314 & 0.723 & 0.294 & 0.310 \\
    XGBoost & 0.621 & 0.481 & 0.860 & 0.284 & 0.493 & 0.354 & 0.672 & 0.376 & 0.412 \\
    XGBoost-C & 0.463 & 0.444 & 0.628 & 0.251 & 0.452 & \textbf{0.257} & 0.501 & 0.322 & 0.411 \\
    
    \midrule
    \textbf{EnCod} &  \textbf{0.315} &  \textbf{0.334} & \textbf{0.349} & \textbf{0.245} & \textbf{0.278} & 0.271  & \textbf{0.377} & \textbf{0.269} & \textbf{0.280} \\
    \bottomrule
    
    \end{tabular}}
    \label{tbl:province}
\end{table*}

\begin{table}[htbp!]
    \centering
    \caption{The NRMSE performance for long-term prediction.}
    \resizebox{\linewidth}{!}{%
    \begin{tabular}{c|c c c c c c}
    \toprule
    \multirow{2}{*}{\textbf{Model}} & \multicolumn{5}{c}{\textbf{Forecasting interval ($n$)}} \\
    
    \cmidrule{2-7}
    
      & \textbf{1} & \textbf{3} & \textbf{5} & \textbf{7} & \textbf{10} & \textbf{14}\\
    
    \midrule
    
    AR & \textbf{0.215} & 0.336 & 0.475 & 0.657 & 0.820 & 0.796 \\
    ARIMA & 0.289 & 0.370 & 0.395 & 0.461 & 0.453 & 0.342\\
    Seq2seq & 0.393 & 0.436 & 0.430 & 0.438 & 0.438 & 0.316\\
    XGBoost & 0.464 & 0.666 & 0.704 & 0.718 & 0.697 & 0.502\\
    XGBoost-C & 0.429 & 0.513 & 0.496 & 0.468 & 0.528 & 0.389\\

    \midrule
    \textbf{EnCod}  & 0.316 &  \textbf{0.309} & \textbf{0.288} & \textbf{0.304} & \textbf{0.335} & \textbf{0.245}\\
    \bottomrule
    
    \end{tabular}}
    \label{tbl:long_term}
    \vspace{-4mm}
\end{table}

We now focus on a single product category, face masks, to examine the regional and long-term forecasting capability. To examine province-level results, we choose nine representative provinces in China considering the geography, the distance from Hubei, and confirmed cases. We train each province-specific model with its own confirmed cases and searching records to predict their citizens' needs.


Table~\ref{tbl:province} shows the comparison of regional forecasting. Even when learning is fine-tuned over province-level data, EnCod still outperforms baselines in most cases. Compared with other provinces, the EnCod model adopted in Hunan, Henan, Sichuan and Zhejiang, which are places nearby Hubei province, delivers relatively low prediction error. It could be explained as the epidemic more influences these areas. Thus the effect of adding COVID-19 statistics is more helpful and leads to better e-commerce behavior predictions.


Next, to test how well EnCod performs for a longer period prediction, we increase the days of a forecast by changing the $n$ value to $\{1,3,5,7,10,14\}$. Table~\ref{tbl:long_term} displays the results from this long-term prediction. The immediate day prediction $n=1$ performs better for the AR model. For $n>1$, EnCod consistently outperforms all baselines by a substantial margin. In contrast, the AR model performs poorly for longer-term prediction, reaching an NRMSE value above 0.5 after a week or longer prediction. The ability to look beyond a week makes the proposed EnCod model practical and applicable to study future epidemics. 

To summarize, all the above results verify that our method can be applied to global and local forecasts and long-term forecasts of the demand for essential goods, which is crucial and meaningful in the pandemic period. Overall, our method can effectively improve forecasting performance, no matter the historical records of time series are sparse or dense, no matter how long the future is predicted, which shows the model's utility and robustness.

\section{Discussion and Conclusion}

The COVID-19 epidemic has exacerbated difficulties in the sufficient supply of essential products to meet the demand and, consequently, influence people's activities on the e-commerce platform. This paper conducts extensive data analysis to investigate how people's online shopping behaviors respond to the epidemic and discover different behavioral patterns. We find out the disruption in the supply of essential goods in this period led to changes in shopping actions (e.g., a positive rank change of search action for face masks and other epidemic-relevant products). Therefore, we incorporate the epidemic development statistics into demand forecasting and present an EnCod model. The model is simple yet effective in forecasting the demand for COVID-19 related items during the epidemics.

The findings of this research have multiple implications. First, the product-level detailed shopping logs will be anonymized and released to the research community and serve as a critical data source to understand market disruptions during a health pandemic. Second, health professionals and e-commerce marketers can utilize the model for predicting surges in the short-term demand for particular goods under risks from an epidemic. Third, policymakers can review the most relevant product goods identified in this research to understand households' needs. Fourth, the encoder-decoder model (EnCod) can be utilized in domains beyond e-commerce (such as trade data) to review the impact of COVID-19 in other sectors.  

In China, the population's proportion using mobile e-commerce is relatively high, and the country has gone through the leading COVID-19 epidemic development during the first quarter of 2020. However, it will be necessary to determine how this analysis can be repeated in other countries where the proportion of using mobile e-commerce may be lower, or COVID-19 is still in progress. In the post-COVID-19 era, inventory planning and pricing of goods will have to be decided based on multiple data sources, including user demands. Moreover, combining other modality of mobile e-commerce~\cite{cao2020beidian, chen2020beidian} to assist demand forecasting, in consideration of cost-effectiveness, would also be an excellent direction to extend this work. As the first to identify the impact of COVID-19 from this perspective of mobile e-commerce, we believe that this study makes an important contribution to the community.

\bibliographystyle{aaai21}

\newpage
\bibliography{reference}

\begin{thebibliography}{25}
\providecommand{\natexlab}[1]{#1}
\providecommand{\url}[1]{\texttt{#1}}
\providecommand{\urlprefix}{URL }
\expandafter\ifx\csname urlstyle\endcsname\relax
  \providecommand{\doi}[1]{doi:\discretionary{}{}{}#1}\else
  \providecommand{\doi}{doi:\discretionary{}{}{}\begingroup
  \urlstyle{rm}\Url}\fi

\bibitem[{Alon et~al.(2020)Alon, Doepke, Olmstead-Rumsey, and
  Tertilt}]{alon2020impact}
Alon, T.~M.; Doepke, M.; Olmstead-Rumsey, J.; and Tertilt, M. 2020.
\newblock The impact of COVID-19 on gender equality.
\newblock Technical report, National Bureau of Economic Research.

\bibitem[{Arora et~al.(2020)Arora, Charm, Grimmelt, Ortega, Robinson, Sexauer,
  Staack, Whitehead, and Yamakawa}]{Mckinsey_global}
Arora, N.; Charm, T.; Grimmelt, A.; Ortega, M.; Robinson, K.; Sexauer, C.;
  Staack, Y.; Whitehead, S.; and Yamakawa, N. 2020.
\newblock A global view of how consumer behavior is changing amid covid-19.
\newblock \urlprefix\url{http://tinyurl.com/y83vtejl}.

\bibitem[{Bahdanau, Cho, and Bengio(2014)}]{bahdanau2014neural}
Bahdanau, D.; Cho, K.; and Bengio, Y. 2014.
\newblock Neural machine translation by jointly learning to align and
  translate.
\newblock \emph{arXiv Preprint arXiv:1409.0473} .

\bibitem[{Baker et~al.(2020)Baker, Bloom, Davis, Kost, Sammon, and
  Viratyosin}]{baker2020unprecedented}
Baker, S.~R.; Bloom, N.; Davis, S.~J.; Kost, K.~J.; Sammon, M.~C.; and
  Viratyosin, T. 2020.
\newblock The unprecedented stock market impact of COVID-19.
\newblock Technical report, National Bureau of Economic Research.

\bibitem[{Cao et~al.(2020)Cao, Chen, Xu, Wang, Xu, Zhang, and
  Li}]{cao2020beidian}
Cao, H.; Chen, Z.; Xu, F.; Wang, T.; Xu, Y.; Zhang, L.; and Li, Y. 2020.
\newblock When Your Friends Become Sellers: An Empirical Study of Social
  Commerce Site Beidian.
\newblock In \emph{Proceedings of the 14th International Conference on Web and
  Social Media}, 83--94.

\bibitem[{Chang and Meyerhoefer(2020)}]{chang2020covid}
Chang, H.-H.; and Meyerhoefer, C. 2020.
\newblock COVID-19 and the Demand for Online Food Shopping Services: Empirical
  Evidence from Taiwan.
\newblock Technical report, National Bureau of Economic Research.

\bibitem[{Chen and Guestrin(2016)}]{chen2016xgboost}
Chen, T.; and Guestrin, C. 2016.
\newblock XGBoost: A scalable tree boosting system.
\newblock In \emph{Proceedings of the 22nd ACM SIGKDD International Conference
  on Knowledge Discovery and Data Mining}, 785--794.

\bibitem[{Chen et~al.(2020)Chen, Cao, Xu, Cheng, Wang, and
  Li}]{chen2020beidian}
Chen, Z.; Cao, H.; Xu, F.; Cheng, M.; Wang, T.; and Li, Y. 2020.
\newblock Understanding the Role of Intermediaries in Online Social
  E-commerces: an Exploratory Study of Beidian.
\newblock In \emph{In Proceedings of the ACM on Human Computer Interaction}.

\bibitem[{Cohen(2020)}]{cohen2020scientists}
Cohen, J. 2020.
\newblock Scientists are racing to model the next moves of a coronavirus that's
  still hard to predict.
\newblock \urlprefix\url{http://doi.org/10.1126/science.abb2161}.

\bibitem[{Contreras et~al.(2003)Contreras, Espinola, Nogales, and
  Conejo}]{contreras2003arima}
Contreras, J.; Espinola, R.; Nogales, F.~J.; and Conejo, A.~J. 2003.
\newblock ARIMA models to predict next-day electricity prices.
\newblock \emph{IEEE Transactions on Power Systems} 18(3): 1014--1020.

\bibitem[{Cornwall(2020)}]{SWarren20}
Cornwall, S. 2020.
\newblock Social scientists scramble to study pandemic, in real time.
\newblock \urlprefix\url{http://tinyurl.com/roave7d}.

\bibitem[{Hoffmann et~al.(2020)Hoffmann, Kleine-Weber, Schroeder, Kr{\"u}ger,
  Herrler, Erichsen, Schiergens, Herrler, Wu, Nitsche, M{\"u}ller, Christian,
  and P{\"o}hlmann}]{hoffmann2020sars}
Hoffmann, M.; Kleine-Weber, H.; Schroeder, S.; Kr{\"u}ger, N.; Herrler, T.;
  Erichsen, S.; Schiergens, T.~S.; Herrler, G.; Wu, N.-H.; Nitsche, A.;
  M{\"u}ller, M.; Christian, D.; and P{\"o}hlmann, S. 2020.
\newblock SARS-CoV-2 cell entry depends on ACE2 and TMPRSS2 and is blocked by a
  clinically proven protease inhibitor.
\newblock \emph{Cell} 181(2).

\bibitem[{Huang et~al.(2020)Huang, Wang, Fan, Zhuo, Sun, and
  Li}]{huang2020understanding}
Huang, J.; Wang, H.; Fan, M.; Zhuo, A.; Sun, Y.; and Li, Y. 2020.
\newblock Understanding the Impact of the COVID-19 Pandemic on
  Transportation-related Behaviors with Human Mobility Data.
\newblock In \emph{Proceedings of the 26th ACM SIGKDD International Conference
  on Knowledge Discovery and Data Mining}, 3443--3450.

\bibitem[{Maier and Brockmann(2020)}]{Maier742}
Maier, B.~F.; and Brockmann, D. 2020.
\newblock Effective containment explains subexponential growth in recent
  confirmed COVID-19 cases in China.
\newblock \emph{Science} 368(6492): 742--746.

\bibitem[{Meltzer, Cox, and Fukuda(1999)}]{meltzer1999economic}
Meltzer, M.~I.; Cox, N.~J.; and Fukuda, K. 1999.
\newblock The economic impact of pandemic influenza in the United States:
  priorities for intervention.
\newblock \emph{Emerging infectious diseases} 5(5): 659.

\bibitem[{Mills(1991)}]{mills1991time}
Mills, T.~C. 1991.
\newblock \emph{Time Series Techniques for Economists}.
\newblock Cambridge University Press.

\bibitem[{Omar, Hoang, and Liu(2016)}]{omar2016hybrid}
Omar, H.; Hoang, V.~H.; and Liu, D.-R. 2016.
\newblock A hybrid neural network model for sales forecasting based on ARIMA
  and search popularity of article titles.
\newblock \emph{Computational Intelligence and Neuroscience} 2016.

\bibitem[{Rocha, Cortez, and Neves(2007)}]{rocha2007evolution}
Rocha, M.; Cortez, P.; and Neves, J. 2007.
\newblock Evolution of neural networks for classification and regression.
\newblock \emph{Neurocomputing} 70(16-18): 2809--2816.

\bibitem[{Schoenbaum(1987)}]{schoenbaum1987economic}
Schoenbaum, S.~C. 1987.
\newblock Economic impact of influenza: the individual's perspective.
\newblock \emph{The American Journal of Medicine} 82(6): 26--30.

\bibitem[{Shen(2015)}]{shen2015analysis}
Shen, C. 2015.
\newblock Analysis of detrended time-lagged cross-correlation between two
  nonstationary time series.
\newblock \emph{Physics Letters A} 379(7): 680--687.

\bibitem[{Sumner et~al.(2020)Sumner, Hoy, Ortiz-Juarez
  et~al.}]{sumner2020estimates}
Sumner, A.; Hoy, C.; Ortiz-Juarez, E.; et~al. 2020.
\newblock Estimates of the Impact of COVID-19 on Global Poverty.
\newblock Technical report, WIDER Working Paper 2020/43.

\bibitem[{Sutskever, Vinyals, and Le(2014)}]{sutskever2014sequence}
Sutskever, I.; Vinyals, O.; and Le, Q.~V. 2014.
\newblock Sequence to sequence learning with neural networks.
\newblock In \emph{Advances in Neural Information Processing Systems},
  3104--3112.

\bibitem[{Tian et~al.(2020)Tian, Liu, Li, Wu, Chen, Kraemer, Li
  et~al.}]{tian2020investigation}
Tian, H.; Liu, Y.; Li, Y.; Wu, C.-H.; Chen, B.; Kraemer, M.~U.; Li, B.; et~al.
  2020.
\newblock An investigation of transmission control measures during the first 50
  days of the COVID-19 epidemic in China.
\newblock \emph{Science} 368(6491): 638--642.

\bibitem[{Wrapp et~al.(2020)Wrapp, Wang, Corbett, Goldsmith, Hsieh, Abiona,
  Graham, and Mclellan}]{wrapp2020cryo-em}
Wrapp, D.; Wang, N.; Corbett, K.~S.; Goldsmith, J.~A.; Hsieh, C.; Abiona, O.;
  Graham, B.~S.; and Mclellan, J.~S. 2020.
\newblock Cryo-EM structure of the 2019-nCoV spike in the prefusion
  conformation.
\newblock \emph{Science} 367(6483): 1260--1263.

\bibitem[{WTO(2020)}]{wto2020ecommerce}
WTO. 2020.
\newblock E-commerce, trade and the COVID-19 pandemic.
\newblock Technical report, World Trade Organization.

\end{thebibliography}

\newpage

\title{Disruption in the Chinese E-Commerce During COVID-19}

\section{Appendix}

\vspace{3mm}
This is the supplementary material for the submission 4638 to AAAI-21. We share the per-product-level logs of browsing, sales, and search logs at \url{https://bit.ly/3kcAEN5}. This data, containing before and after e-commerce logs, will be of great value to the research community to study the impact of a health risk such as COVID-19.

\subsection{List of Top-Products Affected by COVID-19} \label{append:ranking}

The Modeling Market Descriptions section revealed which items saw large disruptions in purchasing during COVID-19. Table~\ref{tab:products_top} and Table~\ref{tab:products_bottom} display the top-20 and bottom-20 products with their highest and lowest relative popularity. As we aggregate data by each week, the week of January 20 will contain events related to both the Chinese new year holiday as well as the COVID-19 lock impact.  

The relative ranking $RP(c,t)$ of COVID-19 related products became very high at the end of January. Especially in the peak period, the corresponding absolute ranking of masks, disinfectants, daily necessities, and hand sanitizers were very high among thousands of items in the Beidian platform, top-1 during the week of January 20, top-2 during the week of February 3, top-1 during the week of January 27, top-8 during the week of January 27, respectively.

\subsection{Evaluation Metrics} \label{append:metrics}
Among the evaluation metrics we employ, here we describe how we compute the Normalized Root Mean Square Error (NRMSE)~\cite{rocha2007evolution}:
\begin{equation}
\begin{split}
    &\mathrm{NRMSE}( y, \hat{y}) = \frac{ \mathrm{RMSE}\left ( y,\hat{y} \right ) }{y_{max} - y_{min}},
\end{split}
\end{equation}
\noindent where $N$ is the number of test samples, $y_i, i\in[1,N]$ represents the ground truth, and $\hat{y}_i,i\in[1,N]$ represents the predicted values.


\subsection{Time-Lagged Cross-Correlation Results}
\label{append:TTLC_behavior_snack}

The Modeling Market Descriptions section showed results on time-lagged cross-correlation (TLCC). The method shifts two sequences relatively in time (time-lagged) and calculates the correlation between them (cross-correlation). Therefore, it can analyze non-stationary time series and quantify how the three shopping actions appear upon a pandemic. We use a rolling window and set each window to be a size of 21 days, and the lagged time is set from 0 to 6 days between the two series. 

\begin{figure}[htpb!]
\begin{minipage}[t]{0.5\textwidth}
\centering
\hspace{-1mm}
\subfigure[Purchasing on snacks]{
{\label{subfig:snack_sales}}
\includegraphics[width=0.99\linewidth]{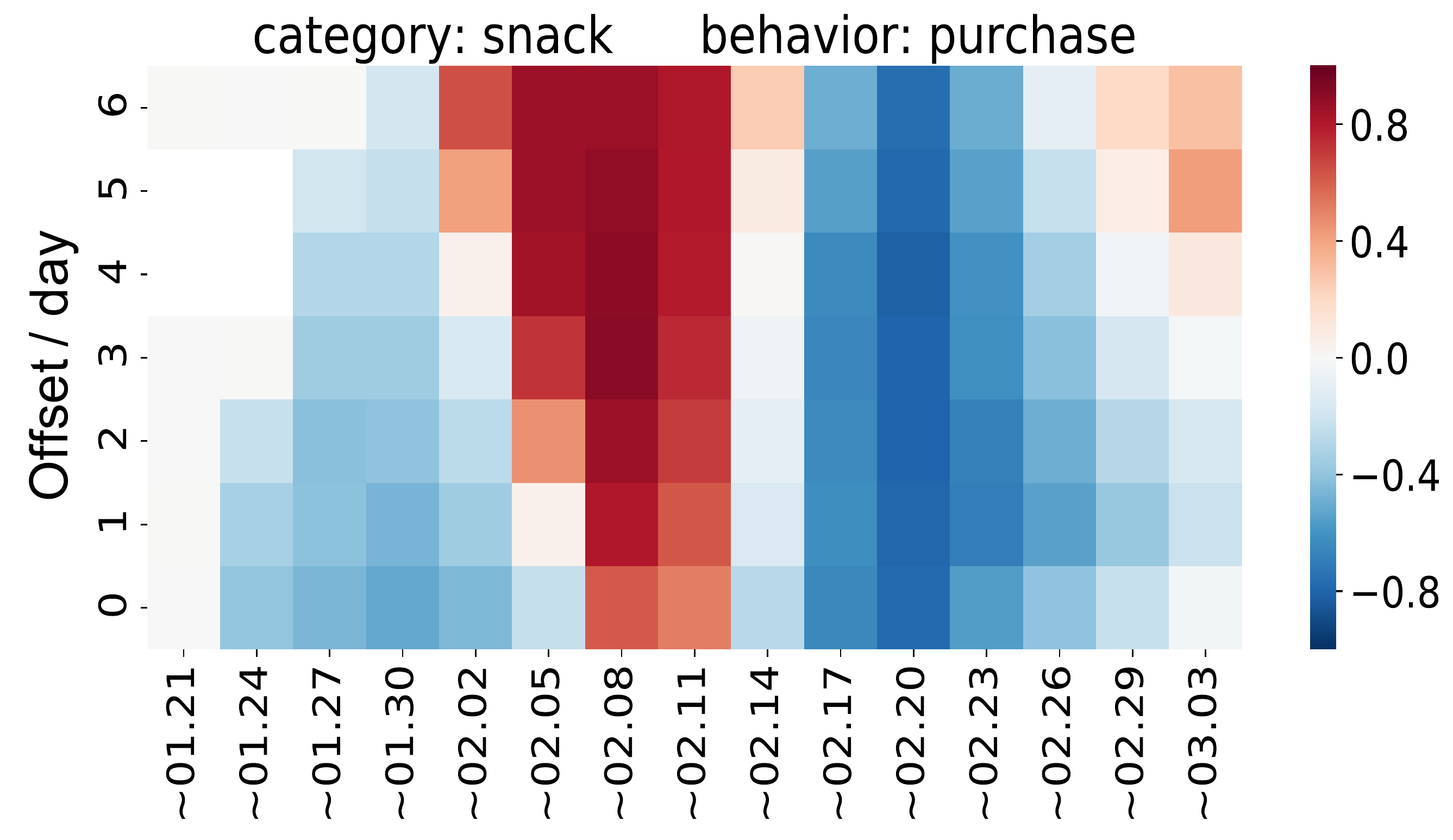}
}
\hspace{-2mm}
\subfigure[Browsing on snacks]{
{\label{subfig:snack_browse}}
\includegraphics[width=0.99\linewidth]{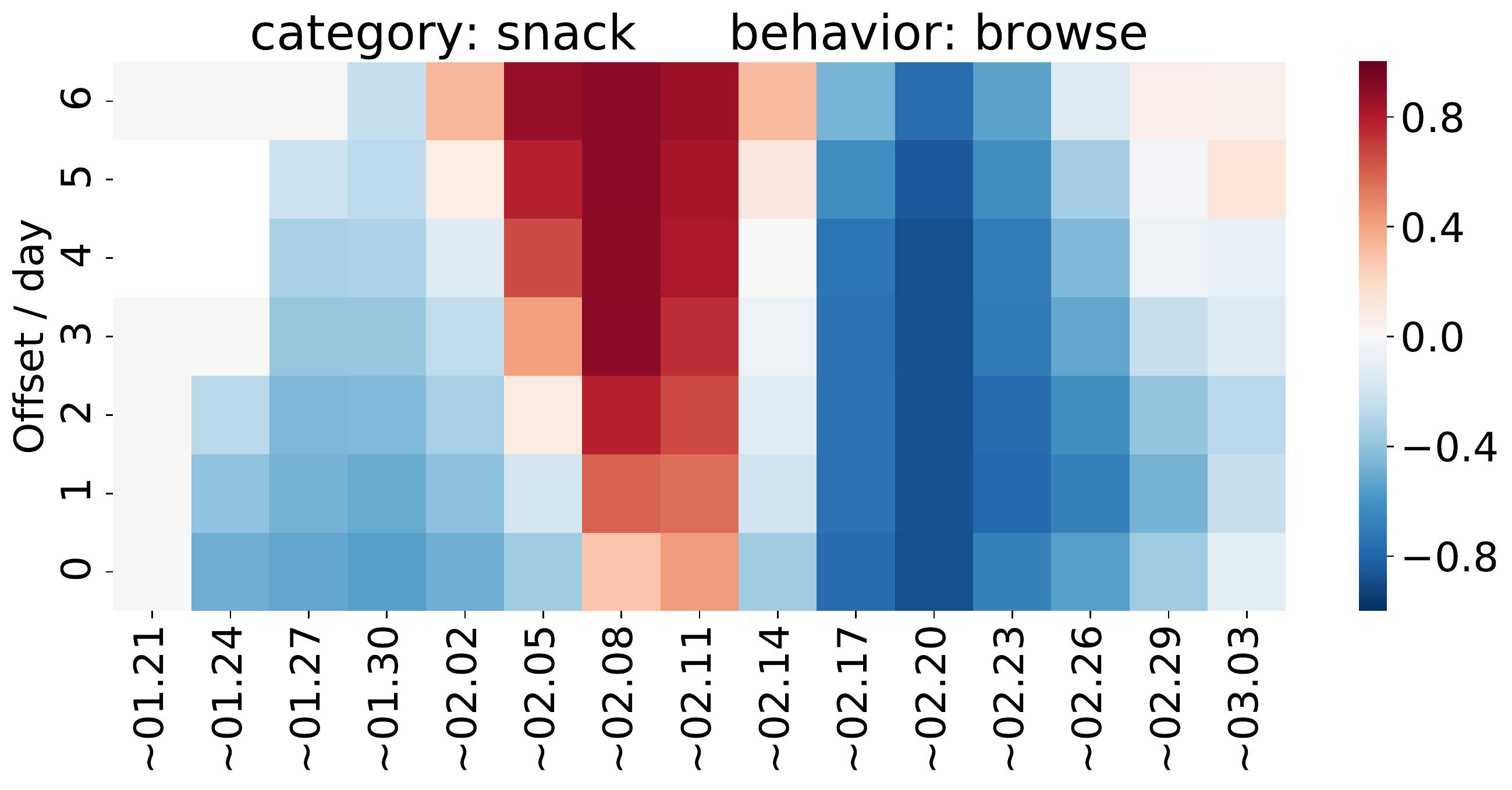}
}
\hspace{-2mm}
\subfigure[Searching on snacks]{
{\label{subfig:snack_search}}
\includegraphics[width=0.99\linewidth]{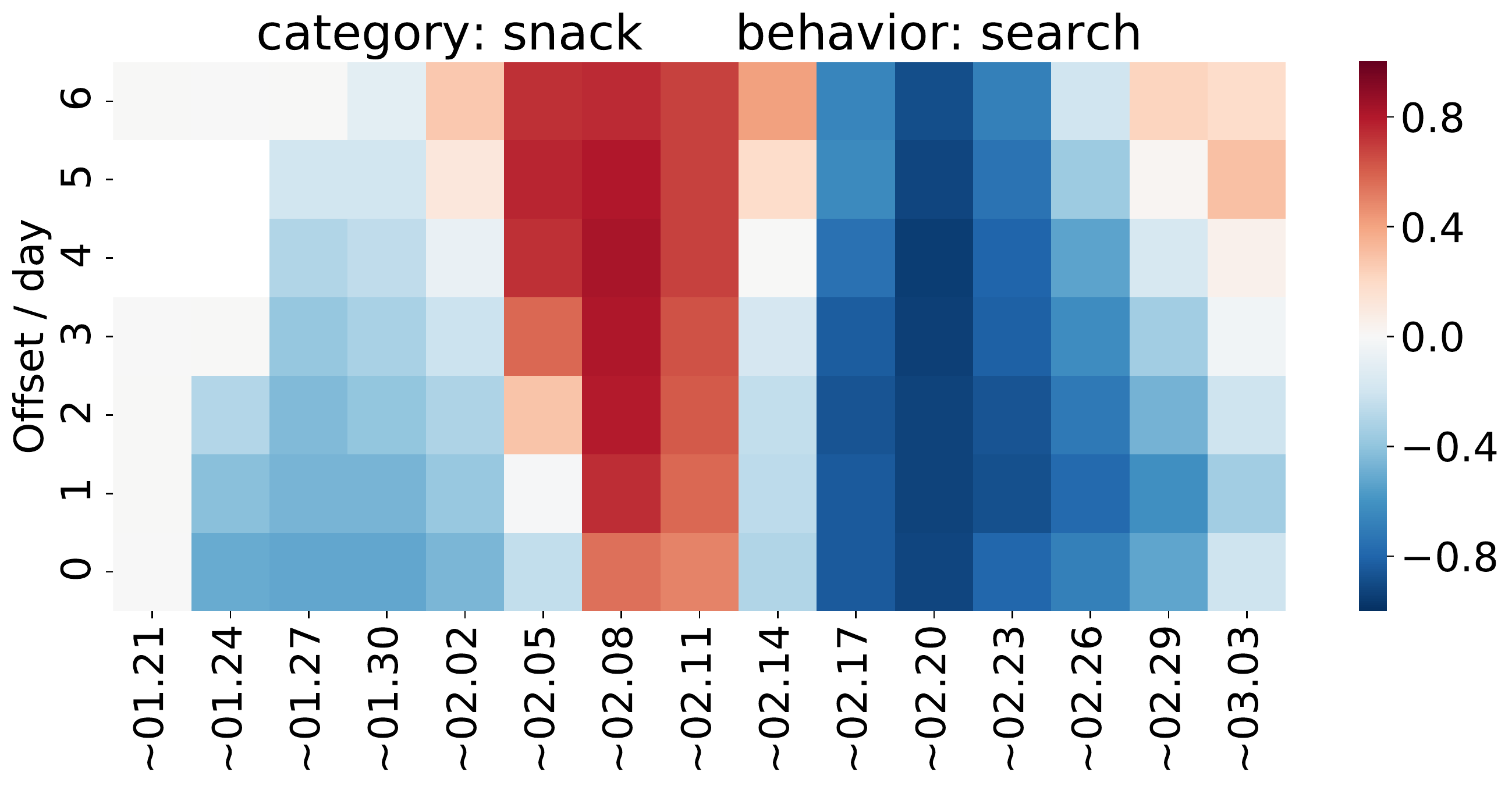}
}
\vspace*{-2mm}
\caption{The TLCC results of searching and browsing behaviors on snacks.}
\label{fig.snack}
\end{minipage}
\end{figure}

The trends seen on browsing and searching are similar to that of purchasing for snacks in Figure~\ref{fig.snack}. Because the supply has nott been affected, the behavioral response during the epidemic is similar across different action types. 

\begin{table*}[!thb]
    \centering
    \caption{Top-20 products with the highest relative popularity. The Peak time column displays the first day of the examined week that reached the largest rank change (i.e. $\arg\min \limits_{t} \mathrm{rank.(t)}$). Note that some of the purchases in the week of January 20th to 26th may be related to the new year's holiday. The city of Wuhan went through lockdown since January 23rd. }
    \resizebox{\linewidth}{!}{%
    \begin{tabular}{l|rrrr|rrr|rrr}
    \toprule
         & \multicolumn{4}{c}{Purchasing} & \multicolumn{3}{|c|}{Browsing} & \multicolumn{3}{c}{Searching} \\ \cmidrule{2-11}
        Product category & $\max \limits_{t} RP(c,t) $ & $\mathrm{rank.}(t_0)$ &  $\mathrm{rank.}$ & Peak time & $\mathrm{rank.}(t_0)$ &  $\mathrm{rank.}$ & Peak time & $\mathrm{rank.}(t_0)$ &  $\mathrm{rank.}$ & Peak time \\
        \midrule
Masks                           & 2.44 & 273 & 1  & Jan 20 & 341 & 1  & Jan 20 & 253  & 1   & Jan 20 \\
Disinfectants                   & 2.35 & 448 & 2  & Feb  3  & 607 & 2  & Feb  3  & 476  & 16  & Feb  10 \\
Face towels                     & 2.20 & 943 & 6  &  Jan 20 & 993 & 14 & Jan 20 & 1234 & 406 &  Jan 20 \\
Jewlery                         & 2.19 & 466 & 3  &  Jan 20 & 398 & 4  & Jan 20 & 402  & 9   &  Jan 20 \\
Accessories                     & 1.95 & 177 & 2  &  Jan 20 & 180 & 2  & Jan 20 & 201  & 3   &  Jan 20 \\
Mango                           & 1.86 & 364 & 5  & Mar 2  & 528 & 14 & Feb  17 & 805  & 76  & Mar 2  \\
Daily necessities               & 1.80 & 63  & 1  &  Jan 27 & 66  & 3  & Jan 27 & 66   & 4   & Feb  3  \\
Hand sanitizers                 & 1.71 & 408 & 8  &  Jan 27 & 565 & 4  & Feb  10 & 427  & 20  & Feb  3  \\
Online courses for children     & 1.70 & 602 & 12 &  Jan 20 & 407 & 59 & Jan 20 & 472  & 307 &  Jan 20 \\
Children masks                  & 1.62 & 784 & 19 &  Jan 27 & 783 & 45 & Jan 27 & 609  & 7   &  Jan 27 \\
Outdoor toys                    & 1.48 & 573 & 19 & Feb  10 & 501 & 25 & Feb  10 & 886  & 574 & Feb  10 \\
Disposable utensils             & 1.36 & 530 & 23 & Feb 3  & 746 & 37 & Feb 3  & 439  & 32  & Feb 3  \\
Tiny bottle sanitizers          & 1.33 & 958 & 45 & Feb 10 & 745 & 84 & Feb 10 & 981  & 972 & Feb 17 \\
Top-up for online entertainment & 1.32 & 710 & 34 &  Jan 20 & 711 & 61 &  Jan 20 & 845  & 258 &  Jan 20 \\
Thermometers                      & 1.30 & 575 & 29 &  Jan 27 & 675 & 31 &  Jan 27 & 534  & 24  & Feb 3  \\
Vitamins                        & 1.22 & 167 & 10 &  Jan 27 & 214 & 13 &  Jan 27 & 205  & 12  &  Jan 27 \\
Wet tissue                      & 1.18 & 46  & 3  & Feb 3  & 117 & 3  & Feb 3  & 84   & 13  & Feb 3  \\
Pineapple                        & 1.16 & 892 & 62 & Feb 24 & 994 & 84 & Feb 24 & 1068 & 110 & Feb 24 \\
Children hats, scarves, gloves      & 1.10 & 316 & 25 & Jan 20 & 251 & 48 & Jan 27 & 132  & 12  & Jan 20 \\
Root vegetables                 & 1.08 & 48  & 4  & Feb 24 & 115 & 12 & Feb 24 & 168  & 24  & Feb 10 \\
        \bottomrule
    \end{tabular}}
    \label{tab:products_top}
\end{table*}

\begin{table*}[!thb]
    \centering
    \resizebox{\linewidth}{!}{%
    \begin{tabular}{l|rrrr|rrr|rrr}
    \toprule
         & \multicolumn{4}{c}{Purchasing} & \multicolumn{3}{|c|}{Browsing} & \multicolumn{3}{c}{Searching} \\ \cmidrule{2-11}
        Product category & $\min \limits_{t} RP(c,t) $ & $\mathrm{rank.}(t_0)$ &  $\mathrm{rank.}$ & Valley time & $\mathrm{rank.}(t_0)$ &  $\mathrm{rank.}$ & Valley  time & $\mathrm{rank.}(t_0)$ &  $\mathrm{rank.}$ & Valley time \\
        \midrule
Flavored milk      & -1.49 & 7   & 215  & Jan 27 & 33  & 266  & Jan 27 & 41  & 131  & Feb 24 \\
Mixed nuts         & -1.45 & 4   & 114  & Feb 3  & 2   & 91   & Feb 3  & 8   & 88   & Feb 24 \\
Cotton clothes     & -1.31 & 32  & 653  & Feb 24 & 6   & 404  & Mar 2  & 4   & 357  & Mar 2  \\
Chocolate          & -1.17 & 11  & 164  & Feb 10 & 25  & 223  & Feb 17 & 37  & 264  & Mar 2  \\
Decorations        & -1.16 & 81  & 1171 & Feb 3  & 203 & 991  & Feb 3  & 356 & 895  & Feb 24 \\
Chinese wine       & -1.08 & 36  & 429  & Feb 17 & 28  & 349  & Feb 17 & 30  & 295  & Mar 2  \\
Warming clothes    & -1.07 & 29  & 341  & Mar 2  & 26  & 264  & Mar 2  & 112 & 441  & Mar 2  \\
Yogurt             & -1.07 & 22  & 260  & Jan 27 & 23  & 286  & Jan 27 & 48  & 164  & Jan 27 \\
Down jackets       & -1.07 & 40  & 467  & Feb 24 & 1   & 155  & Mar 2  & 1   & 101  & Mar 2  \\
Boots              & -1.03 & 44  & 477  & Feb 24 & 5   & 176  & Feb 24 & 3   & 148  & Feb 24 \\
Melon seeds        & -1.02 & 19  & 199  & Jan 20 & 63  & 261  & Jan 27 & 55  & 249  & Jan 20 \\
Snow boots         & -0.94 & 104 & 918  & Feb 24 & 42  & 554  & Feb 24 & 69  & 550  & Mar 2  \\
Red packets        & -0.91 & 172 & 1386 & Mar 2  & 145 & 1377 & Mar 2  & 142 & 1240 & Mar 2  \\
Utensils           & -0.90 & 106 & 844  & Feb 3  & 71  & 858  & Feb 10 & 683 & 960  & Feb 24 \\
Thermal underwear  & -0.84 & 28  & 194  & Feb 24 & 29  & 186  & Mar 2  & 20  & 142  & Mar 2  \\
Traditional snacks & -0.83 & 14  & 94   & Jan 27 & 45  & 120  & Jan 27 & 27  & 96   & Jan 20 \\
Apples             & -0.83 & 3   & 20   & Jan 27 & 21  & 79   & Jan 27 & 61  & 127  & Jan 20 \\
Snow boots for mom & -0.82 & 180 & 1197 & Feb 24 & 99  & 949  & Mar 2  & 64  & 769  & Mar 2  \\
Couplets           & -0.81 & 233 & 1509 & Mar 2  & 184 & 1564 & Mar 2  & 268 & 1538 & Mar 2  \\
Oranges            & -0.76 & 6   & 35   & Mar 2  & 34  & 94   & Jan 27 & 78  & 218  & Jan 27 \\
        \bottomrule
    \end{tabular}}
    \vspace{1mm}
    \caption{Bottom-20 products with the lowest relative popularity. Valley time represents when to reach the lowest ranking (i.e. $\arg\max \limits_{t} \mathrm{rank.(t)}$).}
    \label{tab:products_bottom}
\end{table*}

\end{document}